\documentclass[traditabstract]{aa} 
\usepackage{natbib}
\usepackage{verbatim} 
\usepackage{graphicx}
\usepackage{txfonts}
\begin{document}
\title{Strong  near-infrared emission  in  the sub-AU  disk  of  the
  Herbig~Ae~star HD~163296: evidence of refractory dust?
  \fnmsep\thanks{Based on AMBER observations collected at the VLTI
   (European Southern Observatory, Paranal, Chile) with Arcetri Guaranteed Time program
   081.C-0124, LAOG  Guaranteed Time program 081.C-0794  and open time
   programs 081.C-0851, 081.C-0098.}}  

\author{M.~Benisty  \inst{1,2},  A.~Natta\inst{1},  A.~Isella\inst{3},
  J-P.~Berger\inst{2},    F.~Massi\inst{1},   J-B.~Le~Bouquin\inst{4},
  A.~M\'erand\inst{4},       G.~Duvert\inst{2},      S.~Kraus\inst{5},
  F.~Malbet\inst{2,3},  J.~Olofsson\inst{2},  S.~Robbe-Dubois\inst{6},
  L.~Testi\inst{7}, M.~Vannier\inst{6}, G.~Weigelt\inst{5}} 

\offprints{benisty@arcetri.astro.it}
\institute{INAF-Osservatorio Astrofisico di Arcetri, Largo E.~Fermi~5, 50125
  Firenze, Italy
 \and
 Laboratoire d'Astrophysique de Grenoble, CNRS-UJF UMR 5571, 414 rue de la
 piscine, 38400 St Martin d'H\`eres, France
 \and
 Caltech, MC 249-17, 1200 East California Blvd, Pasadena, CA 91125, USA
 \and
 European Southern Observatory, Casilla 19001, Santiago 19, Chile
 \and
 Max Planck  Institut f\"ur Radioastronomie, Auf dem  H\"ugel 69, 53121
 Bonn, Germany
 \and
 Laboratoire A.~H.~Fizeau, UMR~6525, Universit\'e de Nice-Sophia
 Antipolis, Parc Valrose, 06108 Nice Cedex 02, France
 \and
 European  Southern Observatory,  Karl-Schwarzschild-Strasse  2, 85748
 Garching, Germany\\}
     
\date{Received 2009-07-16; Accepted 2009-11-03}

 
  \abstract
  {We   present   new  long-baseline   spectro-interferometric
    observations of the Herbig~Ae star HD~163296 (MWC~275) obtained in
    the  $H$ and $K$  bands with  the AMBER  instrument at the VLTI.  The
    observations cover a range  of spatial resolutions between $\sim$3
    and  $\sim$12~milliarcseconds,  with  a  spectral  resolution  of
    $\sim$30. With a total of 1481 visibilities and 432 closure phases,
    they represent the most comprehensive $(u,v)$ coverage achieved so
    far for a  young star.  The circumstellar material  is resolved at
    the sub-AU spatial scale 
    and closure  phase measurements  indicate a small  but significant
    deviation  from point-symmetry.  We  discuss the  results assuming
    that the near-infrared excess in  HD~163296 is dominated by the emission
    of a circumstellar disk.  A  successful fit to the spectral energy
    distribution, near-infrared visibilities and closure phases is 
found with a model in which a dominant contribution to the $H$ and
 $K$ band emission originates in an optically thin, smooth and
point-symmetric  region extending  from  about 0.1  to  0.45~AU. At  a
    distance of 0.45~AU from the star, silicates condense, the disk
    becomes optically thick and develops a puffed-up rim, whose skewed
    emission can account for the non-zero closure phases.  
    We discuss the  source of the inner disk  emission and tentatively
    exclude dense molecular gas as well as optically thin atomic or ionized gas as its possible origin. 
    We propose  instead that the  smooth inner emission is  produced by
    very refractory grains in a partially cleared region, extending to
    at least  $\sim$0.5~AU.  If  so, we may  be observing the  disk of
    HD~163296 just before it reaches the transition disk 
    phase. However, we  note that the nature of  the refractory grains
    or, in fact, even the possibility of any grain surviving at the
    very high  temperatures we require  ($\sim 2100-2300$ K  at 0.1~AU
    from the star) is unclear and should be investigated further.}  


   \authorrunning{Benisty et al.}
   \titlerunning{The sub-AU disk of the Herbig~Ae star HD~163296}

   \maketitle
%

\section{Introduction}

Herbig~AeBe stars (HAeBe) are intermediate-mass young stars, surrounded by
large amounts of dust and  gas. The distribution of this circumstellar
material  remains actively  debated.  Various  types  of models  can
reproduce the spectral energy distribution (SED) by considering material in geometrically thin accretion disks
\citep{hillenbrand92},       in       a       spherical       envelope
\citep{miroshnichenko97},       a    puffed-up   inner    disk   rim
\citep{dullemond01,    isella05}   or       a    disk    plus a  halo
\citep{vinkovic06}. Fitting the SED alone is therefore highly ambiguous. 

Near-infrared (NIR) long baseline interferometry has 
allowed us to directly probe the properties of matter within
  the  innermost   astronomical  unit  (AU),   where  key
quantities for the star-disk-protoplanets interactions are set.  The first
interferometric  studies of  HAeBe  showed that  the NIR  characteristic
sizes were larger than expected by classical accretion disk models
\citep{millangabet01}, and were found to be correlated with the stellar
luminosity \citep{monnier02}.  This supports the idea that the NIR emission is
dominated  by  the thermal  emission  of  hot  dust   heated  by  stellar
radiation. \cite{natta01} suggested that an inner, optically thin
  cavity produced by dust sublimation exists inside the disk. At the edge of
this region, where dust condensates, the disk is expected to puff up
  because of the direct illumination from the star
\citep{dullemond01,isella05},  explaining  the  size-luminosity  law
  derived for Herbig~Ae (and late Be) stars by \cite{monnier02}.   
Based  on  a  small  number of  interferometric  observations,  simple
geometrical models  were proposed to explain the  global morphology of
these    regions   \citep{millangabet01,eisner04,monnier05,monnier06}.
However, when  larger sets of  data became available, it  became clear
that the  regions probed by  NIR interferometry are much  more complex
and  that   a  deeper   understanding  requires  the   combination  of
photometric and multi~wavelength interferometric measurements at the
milli-arcsecond resolution and more sophisticated models. 

In this  study, we present an  analysis of the  inner disk surrounding
the HAe star HD~163296  (MWC275). This isolated Herbig~Ae star
  is described well by a spectral type of A1, a $\sim$30~L$_\odot$
luminosity,        and        a       $\sim$2.3~M$_\odot$        mass
  \citep{vandenancker98,natta04,montesinos09}.  It is is located at 
122$^{+17}_{-13}$~pc and  exhibits a  NIR excess interpreted  as the
  emission  from   a  circumstellar  disk   \citep{hillenbrand92}.  A
large-scale disk was detected in scattered light \citep{grady00}, as well as at 
millimeter  wavelengths \citep{mannings97}.   This inclined  disk,
traced  out to  540~AU,  was found  to  be in  Keplerian rotation,  and
probably evolving towards a debris disk phase \citep{isella07}.  In addition,
it  also exhibits  an  asymmetric outflow  on  large scales  ($\geq$27'')
perpendicular  to the  disk, with  a  chain of  six Herbig-Haro  knots
(HH409)     that     traces    the     history     of    mass     loss
\citep{devine00,wassell06}. 
The spectrophotometric  observations that probe  the intermediate and
small spatial scales are also 
compatible with the presence of a disk. \cite{doucet06} studied  the warm  dust emitting in  the mid-infrared,
located in the surface layers  of the intermediate regions of the disk
(30-100~AU) and concluded that the emission was consistent with a disk
that has little flaring.  This conclusion is consistent with the
classification of HD~163296 by  \cite{meeus01} in their Group~II, whose
SED can be explained by assuming that the inner part of the disk shields 
the outer part  from stellar radiation. In the  innermost regions, the
far-UV  emission lines  have  been attributed  to  optically thin  gas
accreting onto the stellar surface, a magnetically confined wind, or
 shocks at the base of the jet \citep{deleuil05,swartz05}.  Weak
  X-ray emission (L$_x$/L$_\star \sim 5 10^{-6}$) was detected on
  large scales  and attributed to  the jet \citep{gunther09}.   In the
NIR, the SED time variability was  interpreted to be due to changes in
the inner disk  structure, on timescales similar to  the generation of
the HH objects \citep{sitko08}.  

With  a disk,  signs  of  accretion, and  a  bipolar outflow,  HD~163296
provides an excellent case study to understand how  circumstellar
material is distributed on the sub-AU scale. Its NIR disk was resolved
by IOTA, PTI, and Keck-Interferometer 
\citep{millangabet01,eisner09, monnier05}, and at milli-arcsecond (mas) resolution with 
the long CHARA baselines \citep{tannirkulam} (T08, hereafter). T08
found  that their observations  could not  be  reproduced using
models where the majority of the $K$ band emission originates in a dust rim, but that an additional NIR emission
inside the dust sublimation  radius could explain the visibilities and
the SED.  They interpreted this additional emission as being produced by 
gas,     as     suggested      for     other     Herbig~AeBe     stars
\citep{eisner07,isella08,kraus08}.  The advent of spectro-interferometry, as provided by the AMBER
instrument at VLTI, allows us to simultaneously measure the emission at
various  NIR  wavelengths  and  consequently,  to  derive  temperature
profiles  for the  emission,  bringing additional  constraints on  its
nature.   In this  paper, we  present  an observational  study of  the
circumstellar disk around HD~163296 at  the sub-AU scale, using the largest
interferometric dataset obtained  for a young star so far.  The paper is
organized    as    follows:     in    Sect.~2,    we    describe    the
spectro-interferometric observations obtained at AMBER/VLTI and the data
processing;  in  Sect.~3,  we  present the  obtained  visibilities  and
closure phases.  In Sect.~4, we outline a successful disk model that reproduces
all observables and we discuss its physical origin.  We summarize our
results in Sect.~5.

\begin{figure}[t]
\centering
\includegraphics[width=0.4\textwidth]{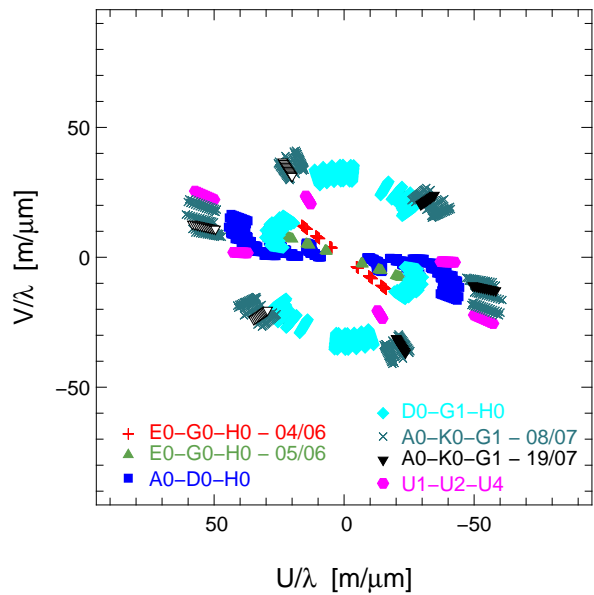}
\caption{\label{uv} $(u,v)$ plane coverage of the observations in spatial
frequencies. The observing nights are plotted with different symbols
  and the  corresponding telescope configurations are  reported in the
  figure.}  
\end{figure}

\begin{table*}[t]
\centering
\caption{\label{obs}       Log       of      the       interferometric
  observations.  The seeing and $\tau_{0}$ were measured at 650~nm.} 
\begin{tabular}{c c c c c c c c c c}
  \hline
  \hline
  Date & Spectral window & Telescope & Baseline & Projected 
  & Position  & Calibrator & Seeing & $\tau_{0}$ & FINITO \\
 &     [$\mu$m]     &    configuration&   name&   length  B$_{\rm{L}}$~[m]   &
  angle B$_{\rm{PA}}$~[$^\circ$] & name & [''] & [ms] & \\
  \hline 
  \hline 
24-05-2008 &[1.67-1.80]-[2.06-2.42] & A0-D0-H0 & A0-D0 & 26.7 & 81.6 &
 HD156897 & 1.0 & 3.9 & Y \\
 & & & D0-H0 & 53.5 & 81.6 & & & &  \\
 & & & A0-H0 & 80.2 & 81.6 & & & &  \\
 26-05-2008  & [1.62-1.79]-[2.05-2.31]  & D0-G1-H0  & D0-G1  &  69.7 &
 138.2& HD156897& 0.8& 4.4 & Y \\
 04-06-2008 & [1.60-1.82]-[2.03-2.33] &  E0-G0-H0 & E0-G0 & 14.3 &56.5
 & HD156897 & 0.6& 3.9 & N\\
 & & & H0-G0 & 28.7 & 56.5& & & & \\
 & & & E0-H0 & 43.0 & 56.5 & & & &\\
 05-06-2008 & [1.61-1.82]-[2.02-2.29] &  E0-G0-H0 & E0-G0 & 15.9 &71.1
 & HD156897& 0.8&4.1 & N\\
  & & & H0-G0 & 31.7 & 71.1& HD163955& & & \\
 & & & E0-H0 & 47.7 & 71.1& & & & \\
24-06-2008 &[1.65-1.80]-[2.02-2.33]  & U1-U2-U4 & U1-U2  & 55.8 &
 33.7 & HD156897 & 1.1& 1.4 & N\\
  & & & U2-U4 & 82.6 & 89.8 & & & &\\
  & & & U1-U4 & 122.8 & 67.6 & & & & \\
06-07-2008 & [1.63-1.79]-[2.01-2.45] & D0-G1-H0 & D0-H0 & 60.2 & 70.4&
 HD160915 & 1.0& 2.3 & N\\
  & & & D0-G1 & 69.2 & 137.1& & & & \\
 & & & G1-H0 & 71.4 & 7.6 & & & & \\
 08-07-2008  &  [1.60-1.82]-[2.01-2.44]& A0-K0-G1  &  A0-G1  & 81.4  &
 121.6& HD160915& 0.8& 2.9 & N\\
  & & & G1-K0 & 86.5 & 29.9& & & & \\
  & & & A0-K0 & 108.4 & 74.9& & & & \\
 19-07-2007  & [1.61-1.80]-[2.06-2.42]  & A0-K0-G1  & A0-G1  &  82.4 &
 124.6& HD156897& 1.0 & 1.6 & N \\
 & & & G1-K0 & 88.6 & 32.7& & & &\\
  & & & A0-K0 & 115.2 & 77.6& & & &\\
 \hline
   \hline    
\end{tabular}
\end{table*}

\section{Observations and data reduction}

\subsection{Observations at VLTI}
HD~163296 was observed in the NIR with the AMBER instrument
\citep{petrov07},   at  the   Very   Large  Telescope   Interferometer
\citep[VLTI;][]{vlti1}, located  at Cerro Paranal,  Chile and operated
by the European Southern Observatory (ESO). The AMBER instrument allows the
simultaneous  combination  of  three  beams  in  the $H$  and  $K$  bands
(\textit{i.e.},  from  1.6 to  2.5~$\mu$m) with spatial  filtering.  The
instrument delivers spectrally  dispersed  interferometric observables
(\textit{e.g.},~visibilities,  closure phases, differential  phases) at
spectral resolutions of up to 12000.  

In the following, we present observations taken at the low spectral
resolution mode  (R$\sim$30) with  the 1.8~m Auxiliary  Telescopes (AT)
and the 8.2~m Unit Telescopes (UTs). 
The data were obtained
within  programs  of both guaranteed  time  and  open  time  observations
(081.C-0794;   081.C-0098;  081.C-0124;  081.C-0851).    HD~163296  was
observed with 14 different baselines of 5 VLTI telescope configurations,
during  8~nights from  May  to  July 2008.   The  longest baseline  is
$\sim$128~m    corresponding    to     a    maximum    resolution    of
3.5~mas in 
the $K$~band,  and of 2.7~mas in the  $H$~band. In this paper,  we use the
VLTI nomenclature to identify the different configurations. A summary
of  the observations can  be found  in Table~\ref{obs},  including the
weather conditions, the average baseline position angles (B$_{\rm{PA}}$), and 
projected lengths (B$_{\rm{L}}$).   The projected baseline is obtained
when the vector between the two telescopes is projected onto the plane
of the sky.  Because of the Earth rotation, measurements with the same
physical baseline but at different hour angles correspond to different
projected baselines.\\ 
All  the  observations were  performed  with  three telescopes,  except
during the night of the 2008 May 26 when only two were available. In
addition  to  HD~163296, three  calibrator  stars (HD~156897,  HD~160915,
HD~163955)  were observed  before  and after  each  measurement on  the
scientific target to correct for instrumental effects.  Their
stellar parameters, including their diameters, can be found in
Table~\ref{calparam}.  About  25\% of the  observations were performed
using the VLTI fringe-tracker FINITO that uses 70\% of the $H$ band
flux to measure the relative optical path difference between the light beams
\citep{lebouquin08}.  

\begin{table}[thb]
  \caption{Star  and  calibrator properties.   The  latter have  been
    chosen                        with                       SearchCal
    ($\textrm{http://www.jmmc.fr/searchcal\_page.htm}$) and getCal ($\textrm{http://mscweb.ipac.caltech.edu/gcWeb/gcWeb.jsp}$) } 
\label{calparam}
  \begin{tabular}{c c c c c c}
    \hline
    \hline
    Star & $V$ &$K$ & $H$ & Spectral Type & Diameter [mas] \\
    \hline 
    HD~163296 & 6.9& 4.8 & 5.5 & A1Ve & / \\
    HD~156897 & 4.4 & 3.1 & 3.1 & F2 & 0.8$\pm$0.2\\
    HD~160915 & 4.9 & 3.8 & 3.9 & F5V & 0.7$\pm$0.1\\
    HD~163955 & 4.7 & 4.5 & 4.6 & B9V & 0.3$\pm$0.2\\
    \hline
    \hline    
  \end{tabular}
\end{table}

\subsection{Photometry}
In  addition   to  this  interferometric  dataset,  we
collected  photometric    data    from   the    literature
\citep{tannirkulam,sitko08}.

\subsection{Data reduction}
The   interferometric   data   reduction   was   performed   following
\cite{tatulli07}, using the  \texttt{amdlib} package (release 2.1) and
the  \texttt{yorick}  interface provided  by  the Jean-Marie  Mariotti
Center (JMMC).  This 
led to spectrally dispersed raw visibilities and closure phases for
all exposures of each  observing file. Not all exposures turned
out to provide useful data. In several cases, instrumental jitter,
insufficient fringe  tracking, and unsatisfactory  light injection into
the instrument led to low contrast interferograms of our rather
faint source, and we had to select the good exposures.
Various  selection  thresholds  were  examined  based  on  the  fringe
signal-to-noise  ratio (SNR) criterion  and led  to the  same absolute
values for the interferometric  observables.  On the other hand, their
accuracy 
changes with varying selections, and the optimal case (\textit{i.e.}, with
the smallest errors) was obtained  with a 20\% and 80\% best exposure
selections  for  the  squared  visibilities and  the  closure  phases,
respectively.  In addition, data obtained at very high airmass with 
unstable fringe tracking were removed.  No selection based on the
optical  path difference (\textit{i.e.},  piston) was  performed, since
the numbers of  useful exposures could have been  a possible source of
bias.  \\ 
For each night, special care was given to the calibration and
stability of the AMBER+VLTI 
instrumental transfer function throughout the whole observing period. 
Measurements of HD~163296 were encircled with observations
of targets of known diameters (see Table~\ref{calparam})
and the  transfer function was interpolated along  all calibrations
of the night.  The errors in the calibrated spectral visibilities and closure phases include the
statistical errors obtained when  averaging the individual exposures as
well as the  errors in the calibration star  diameters. To account for
the variation in the transfer function with time, we 
quadratically added the dispersion over the calibrator measurements to
the errors. The latter is the effect that dominates the error budget.  

With 8 to 14 spectral channels in the $K$~band and 5 to 8 in the
$H$~band, the total data set (after processing) consists of 1000 and 304
spectrally dispersed $K$ band visibilities and closure phases,
respectively, and 481 and 128 $H$ band visibilities and closure
phases,  respectively.   The   processed  data  will  be  made
available for the community in the OI-FITS format \citep{pauls05} on
the                                                             OLBIN
website\footnotemark{}\footnotetext{$\textrm{http://olbin.jpl.nasa.gov}$} 
  in January 2010.  




\begin{figure*}[t]
\centering
\begin{tabular}{cc}
\includegraphics[width=0.44\textwidth]{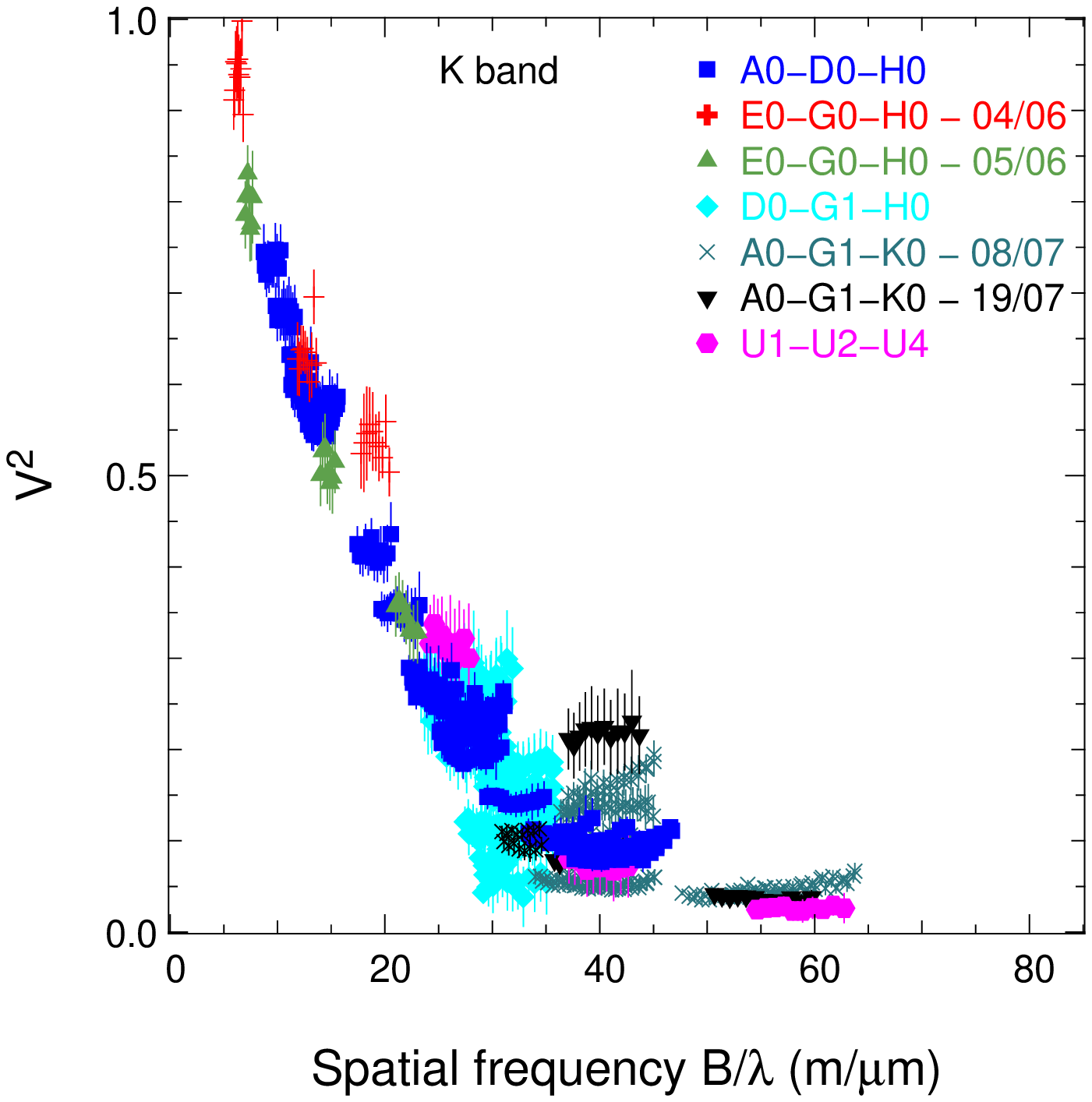}
& 
\includegraphics[width=0.44\textwidth]{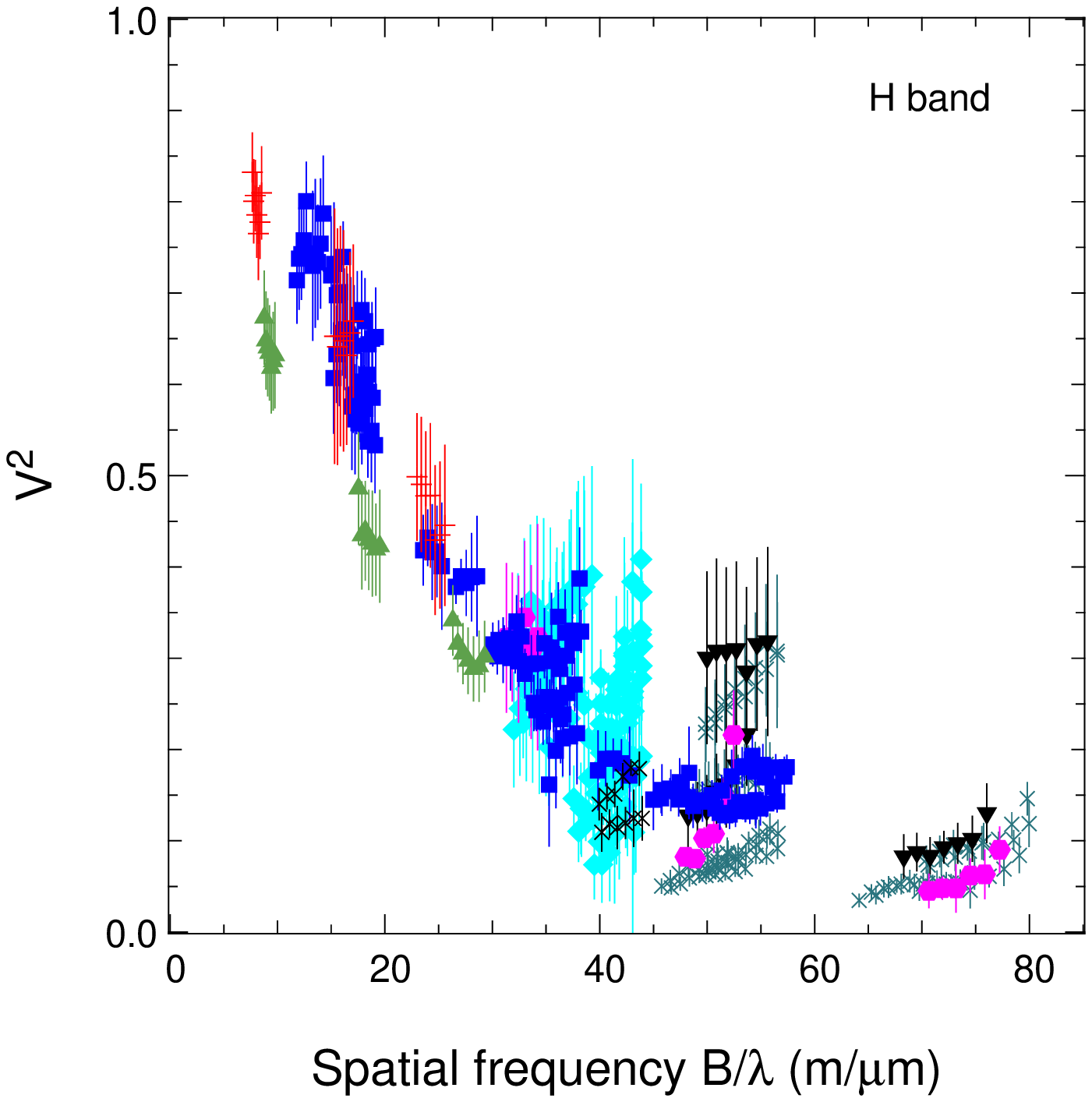}
\end{tabular}
\caption{\label{vis2all}   Squared    visibilities   against   spatial
  frequencies B/$\lambda$ in the $K$ band (left panel) and the $H$ band (right
panel).  Different  symbols correspond to  different configurations,
  that are reported in the upper right corner of the left panel.} 
\end{figure*}

\begin{figure*}[t]
\centering
\begin{tabular}{cc}
\includegraphics[width=0.44\textwidth]{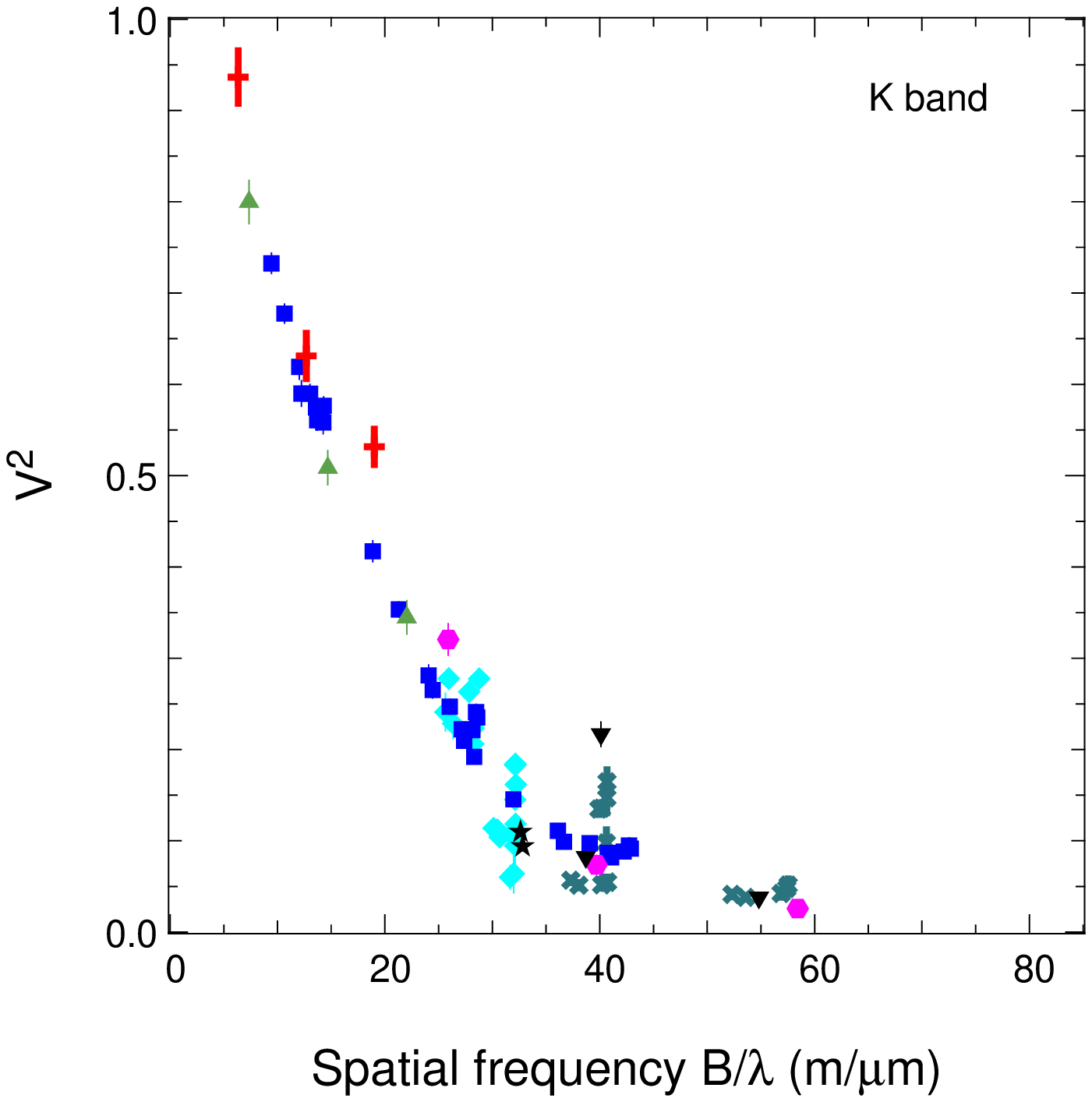}
& 
\includegraphics[width=0.44\textwidth]{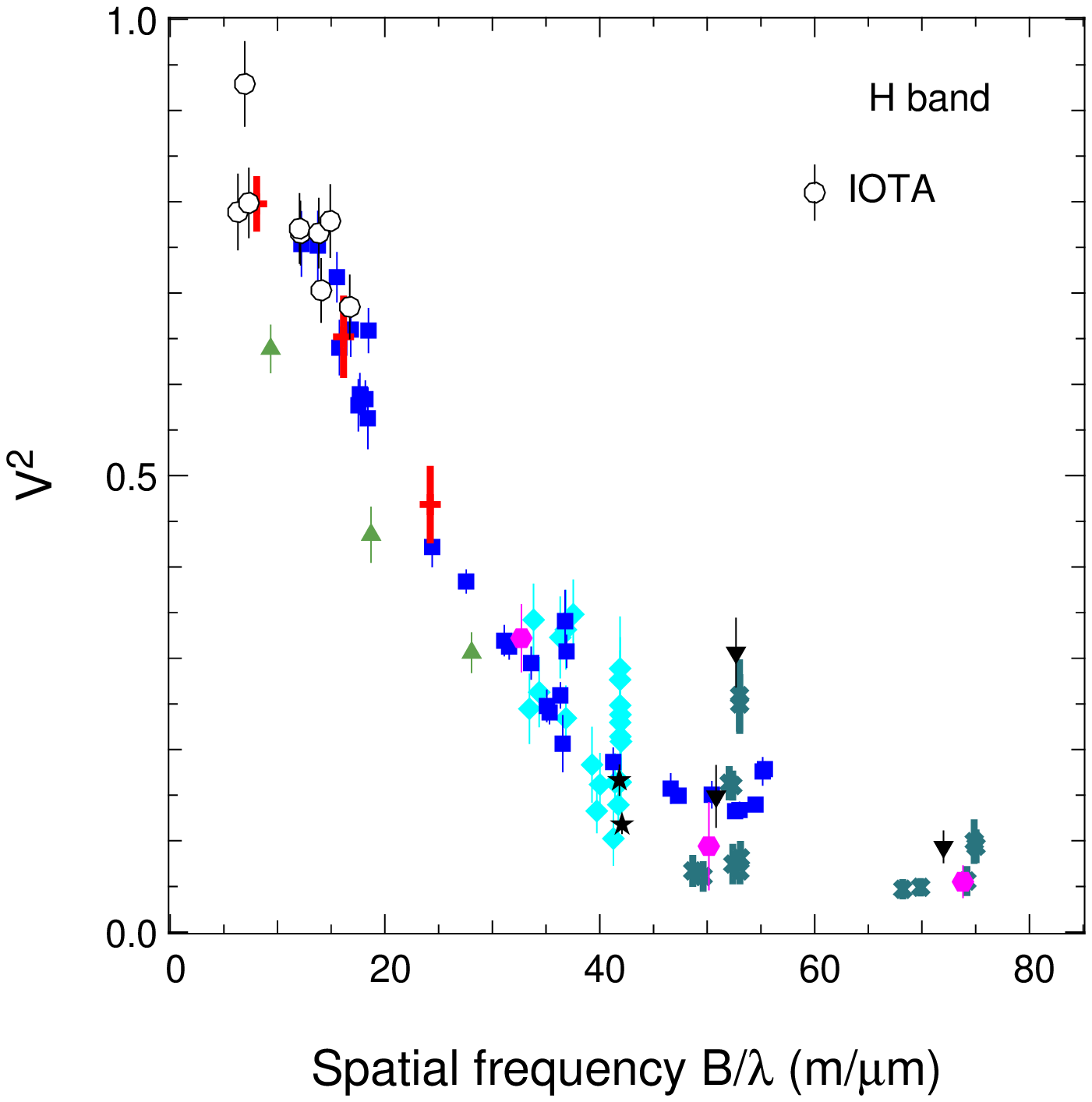}
\end{tabular}
\caption{\label{vis2avg}  Broad-band  squared  visibilities  against
  spatial frequencies B/$\lambda$ in the $K$ band (left panel) and the
  $H$ band (right panel). Different symbols and colors correspond to
  different configurations  that are given  in Fig.~\ref{vis2all}. The
  IOTA points are overplotted in the right panel (empty circles).} 
\end{figure*}

\section{Results}
The set of interferometric data presented in this paper is by far the largest for a
single  pre-main  sequence  star.   The  corresponding  $(u,v)$  plane
coverage is
shown in Fig.~\ref{uv}.  In this section, we present a summary of the results and describe the
main characteristics of the inner region of HD~163296 as measured by NIR interferometry.  For the sake of clarity, we
separate the $H$ and $K$ band results in most of the figures below. 

\subsection{Visibilities}
\label{sec:broadbandvis}
Figure~\ref{vis2all} presents the spectral visibilities as a function of the
spatial frequency (\textit{i.e.},  the  ratio of the projected baseline
length  to the wavelength  of the  observation).  Figure~\ref{vis2avg}
similarly shows 
broad-band visibilities,  obtained when  averaging over  all spectral
channels, for the  $K$ band (left) and the $H$  band (right).  One can
immediately see 
from the figures that  at these spatial resolutions, the circumstellar
matter around 
HD~163296  is   resolved.   Within  the  error  bars,   the  $H$  band
visibilities vary with baseline qualitatively in the same wau as those
of the $K$ band. 
However, at spatial frequencies  higher than 20~m/$\mu$m, the $H$ band
V$^{2}$ are at least 25\% higher than those in the $K$ band. 

The errors in individual points vary significantly from night to night.
However, some of the scatter is not caused by the uncertainties in individual
measurements,  
since  different  observations were  performed  with  baselines of  similar
projected baselines (equivalently, at similar spatial frequencies) but
different position angles, therefore sampling distinct 
orientations  on the  sky.  This  effect means  that the  source
geometry  deviates from  a circular  one, and  can instead  present an
elongated shape, as expected for an inclined disk. 

We  note that in  the $H$  band, at  the shortest  spatial frequencies
(below 10 m/$\mu$m),  the AMBER visibilities do not  reach unity. This
can  be   due  to  flux  from   an  extended  halo   as  suggested  by
\cite{monnier06} from their IOTA $H$ band 
measurements.  All the IOTA points, except one, are consistent with our
measurements (see Fig.~\ref{vis2avg}, right).  On the other hand, this
effect is not clearly seen in the $K$ band, where visibilities are close to
unity at short baselines (\textit{e.g.},  $V^{2}$=0.94$\pm$0.03 at a 
13.8~m baseline).  

\begin{figure}[t] 
\centering
\includegraphics[width=0.45\textwidth]{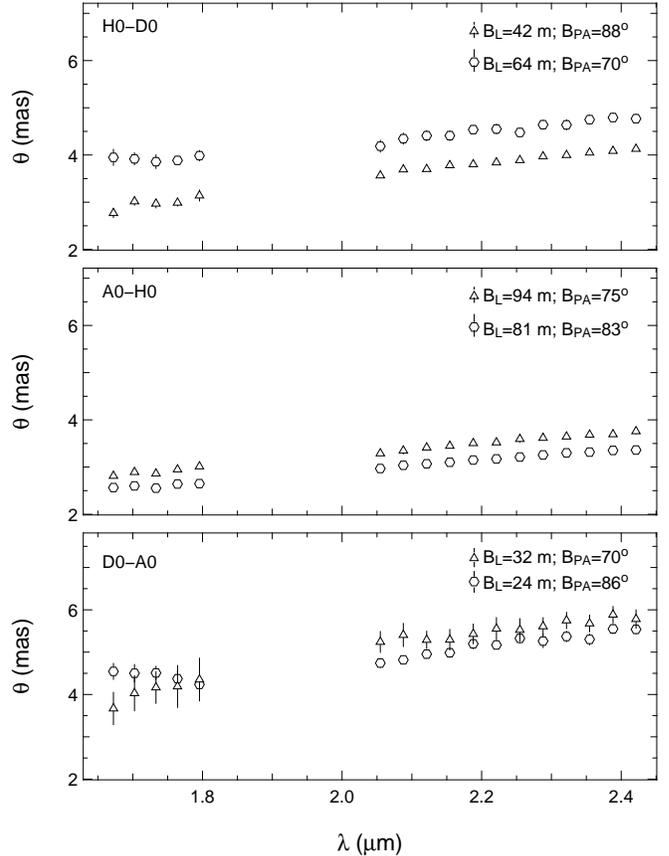}
\caption{\label{lambda} Example of wavelength dependence for the squared visibilities
  obtained on the three  baselines during observations with the A0-D0-H0
  configuration.  In  each  panel,  the  characteristic  size  of  the
  emission is given  at each wavelength across the  $H$ and $K$ band.
  These sizes are  derived from a ring model  of uniform brightness to account for the  change of resolution with wavelength.  The
  triangles show  the maximum  chromaticity over the  whole wavelength
  range,  while  the circles  give  the  minimum  variation.  For  each
  measurement,  the   corresponding  projected  baseline   length  and
  position angle are indicated in the top right corner. } 
\end{figure}

The  observations   show  that   $V^2$  also  depends   on  wavelength
(Fig.~\ref{vis2all}).  This can be related to the physical extension
of the emitting region at various wavelengths and the existence of
temperature gradients within  it.  Since the interferometer resolution
also changes with wavelength, it is natural to visualize this
dependence using  a geometrical model  to convert the  measurements in
angular sizes while taking this effect into account. With this aim, we 
fit the V$^{2}$ in each spectral channel for each measurement, using a
ring of uniform brightness (with a 20\% thickness). 
Although the circumstellar material is 
mainly responsible for the emission in the $H$ and $K$ bands, the star also
contributes to the measured fluxes and visibilities.  
We  estimate a  contribution  from  the  stellar photosphere  of
  $\sim$33\% in H and $\sim$14\% in K (see Sect.~\ref{sec:stellar} for
  details).  The flux from the circumstellar matter was then computed for each AMBER channel 
as  the  difference  between  the  observed NIR  flux  (T08)  and  the
photospheric  flux.   Figure~\ref{lambda}  gives  an  example  of  the
wavelength dependence of the size obtained on the three baselines of the
A0-D0-H0 configuration, in the two extremes cases (observed maximum and minimum 
variations with wavelength).  

For the majority of the measurements (except those obtained at very
small  baselines with  the E0-G0-H0  configuration), the  size  of the
circumstellar matter slightly  increases with wavelength.  To quantify
the wavelength dependence of  the visibility, we studied its variation
over only the $K$ band, $H$ band, and over both bands together. 
Across the $K$ and $H$ band separately, 24\% and 31\%, respectively, of
the measurements show a chromaticity above the 2-$\sigma$ level. 
Over the entire spectral range, a  stronger effect is  expected due to
the  greater  wavelength  interval,  62\%  of the  data  show  a
chromaticity greater than the 2-$\sigma$ level.  This is
consistent  with  the  case  shown  in  Fig.~\ref{lambda},  where  the
characteristic size  of the emitting region increases  between the $H$
and  $K$  bands.   This   trend  is  also  observed  in  the
  spectrally dispersed measurements  across the Br$_{\gamma}$ emission
  line obtained with the Keck-Interferometer \citep{eisner09}.


\begin{figure*}[t]
\centering
\begin{tabular}{ccc}
\includegraphics[width=0.33\textwidth]{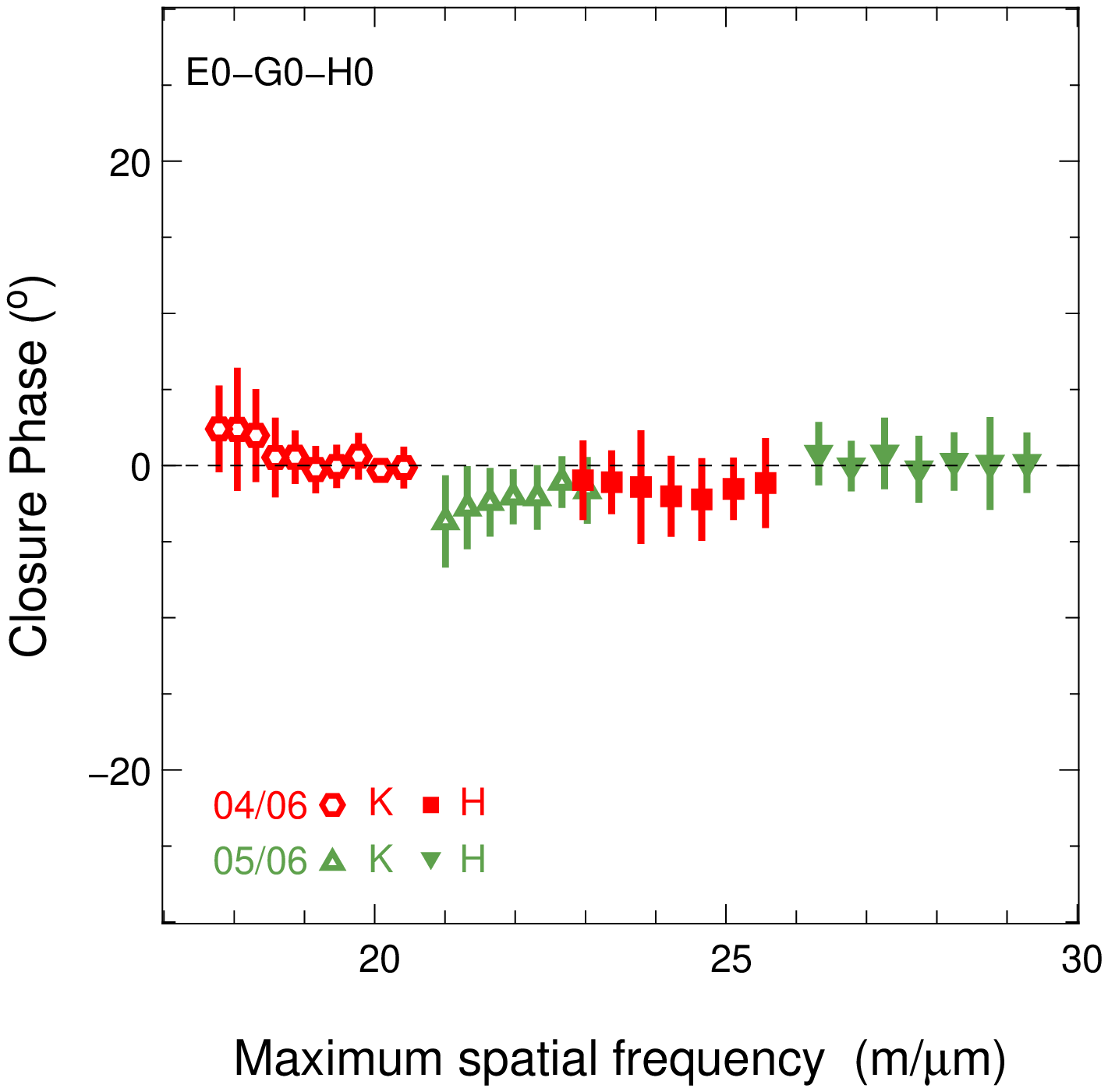}
&
\includegraphics[width=0.33\textwidth]{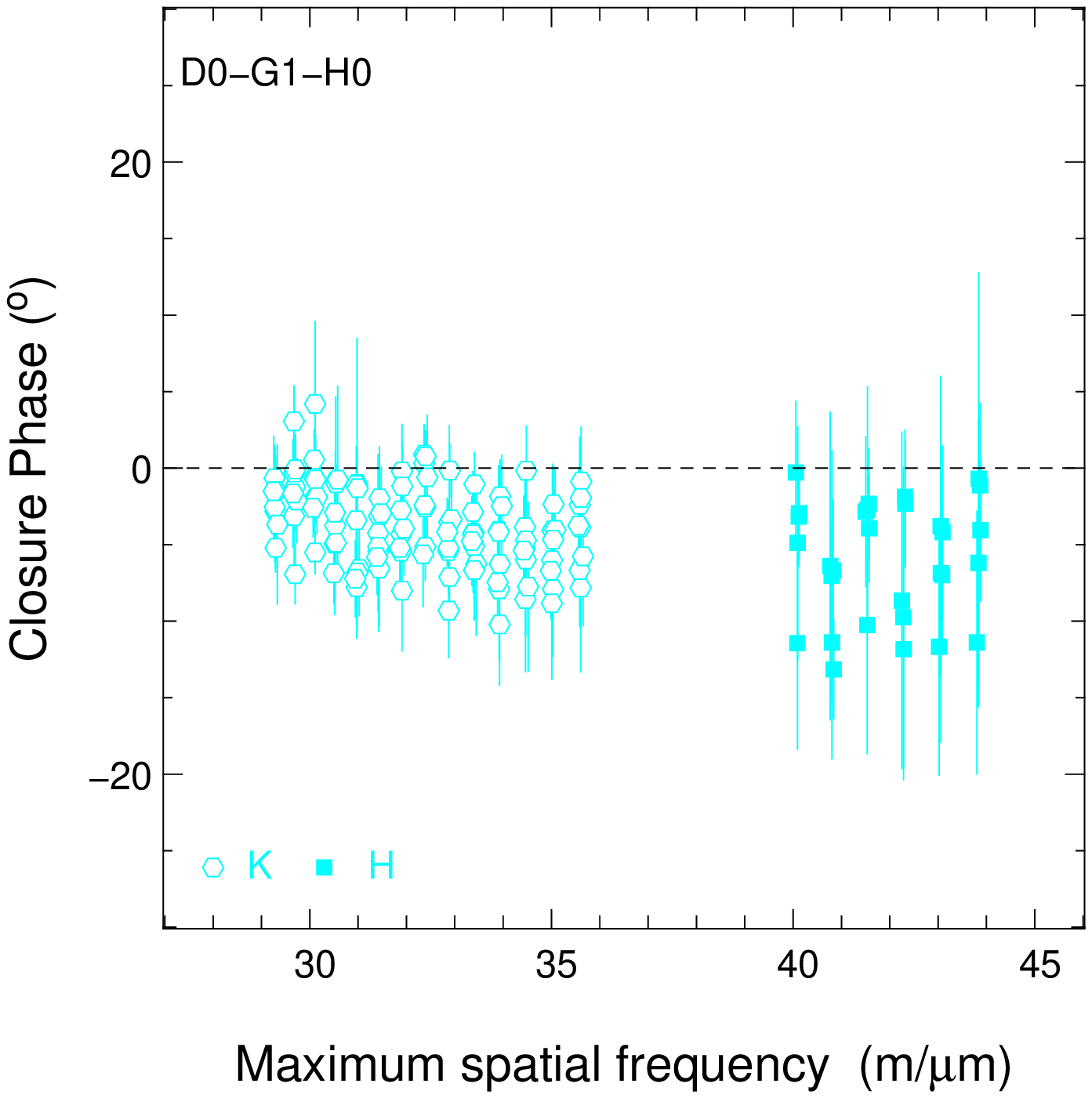}
& 
\includegraphics[width=0.33\textwidth]{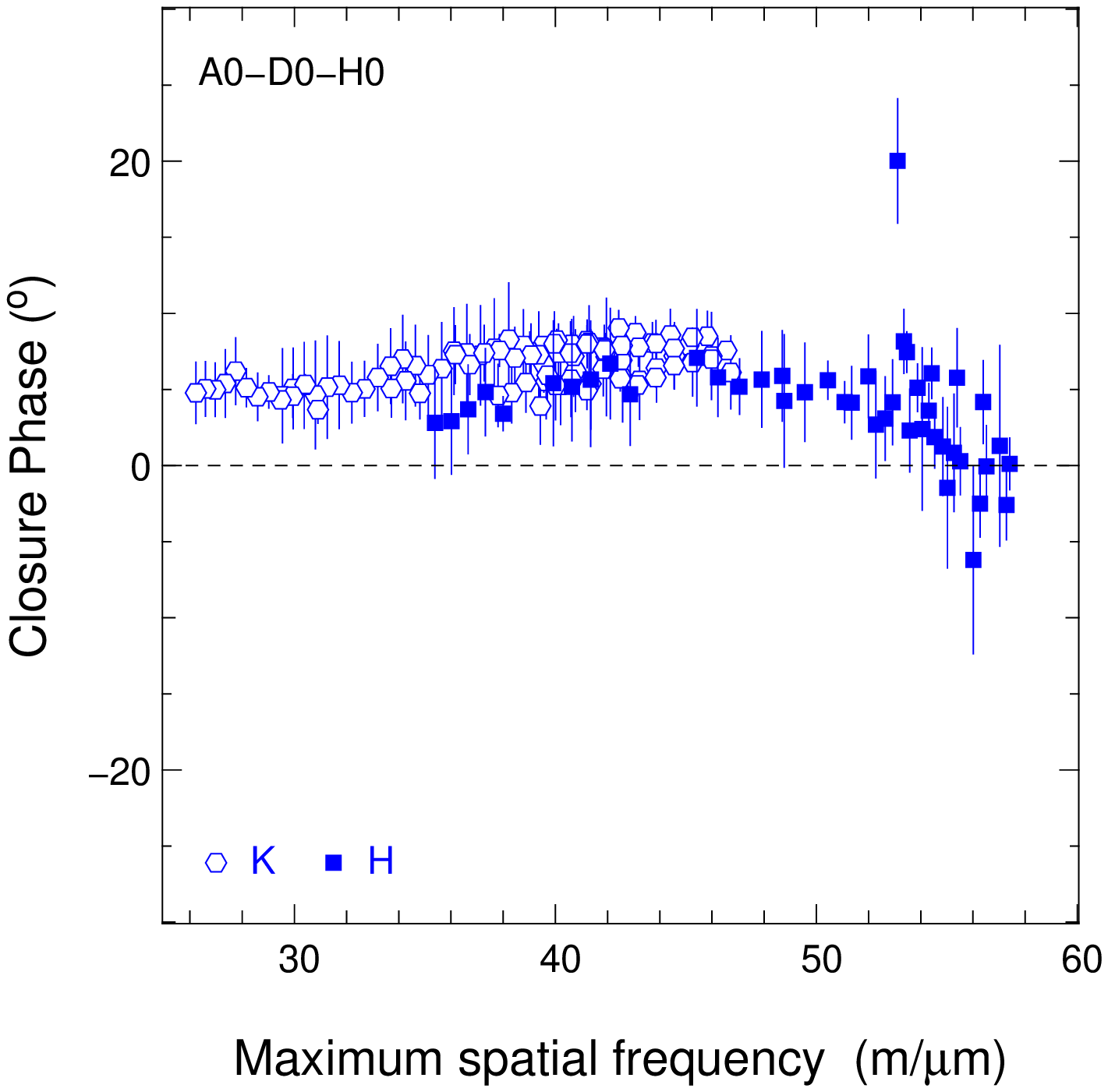}\\
\includegraphics[width=0.32\textwidth]{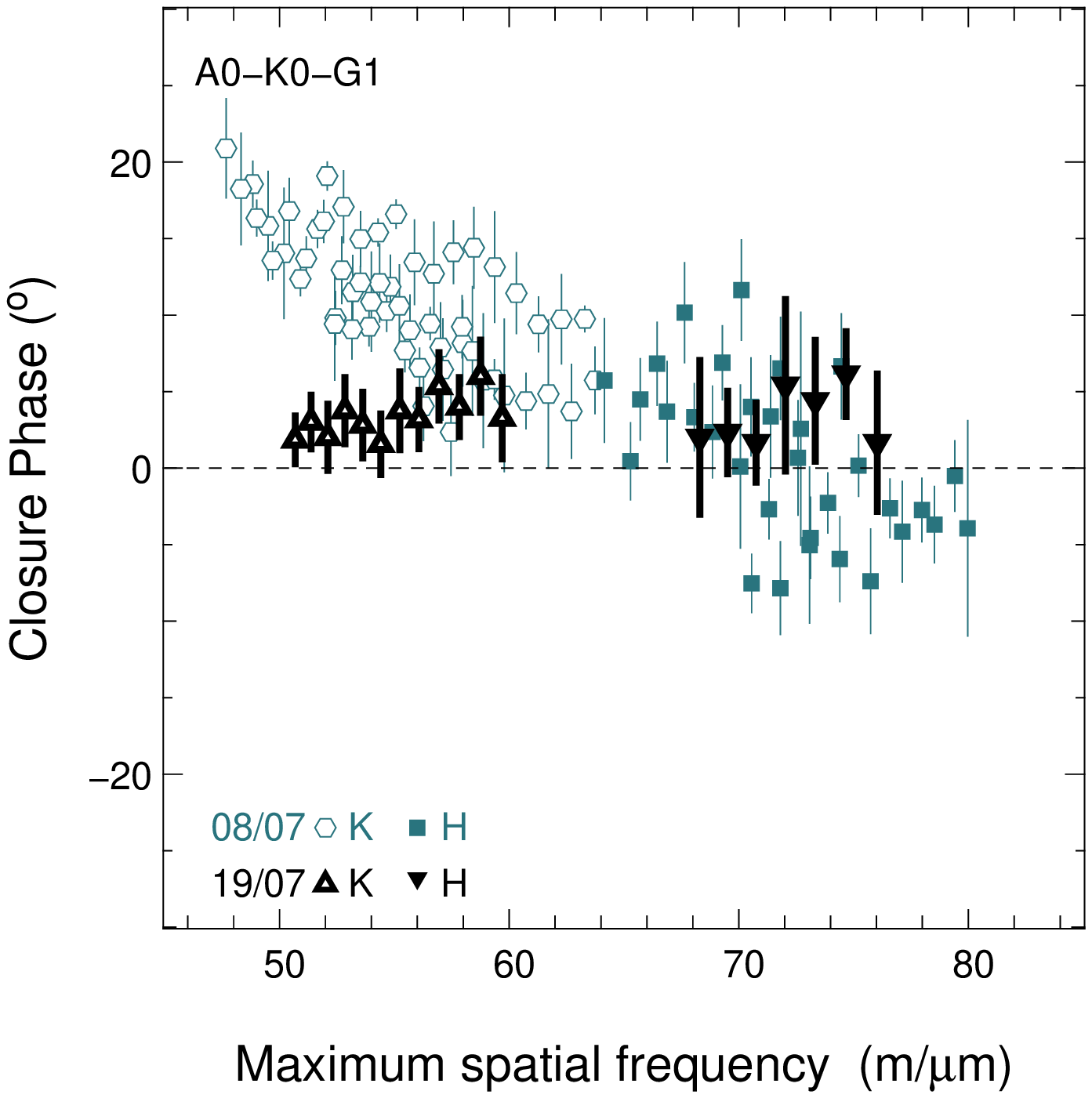}& 
\includegraphics[width=0.32\textwidth]{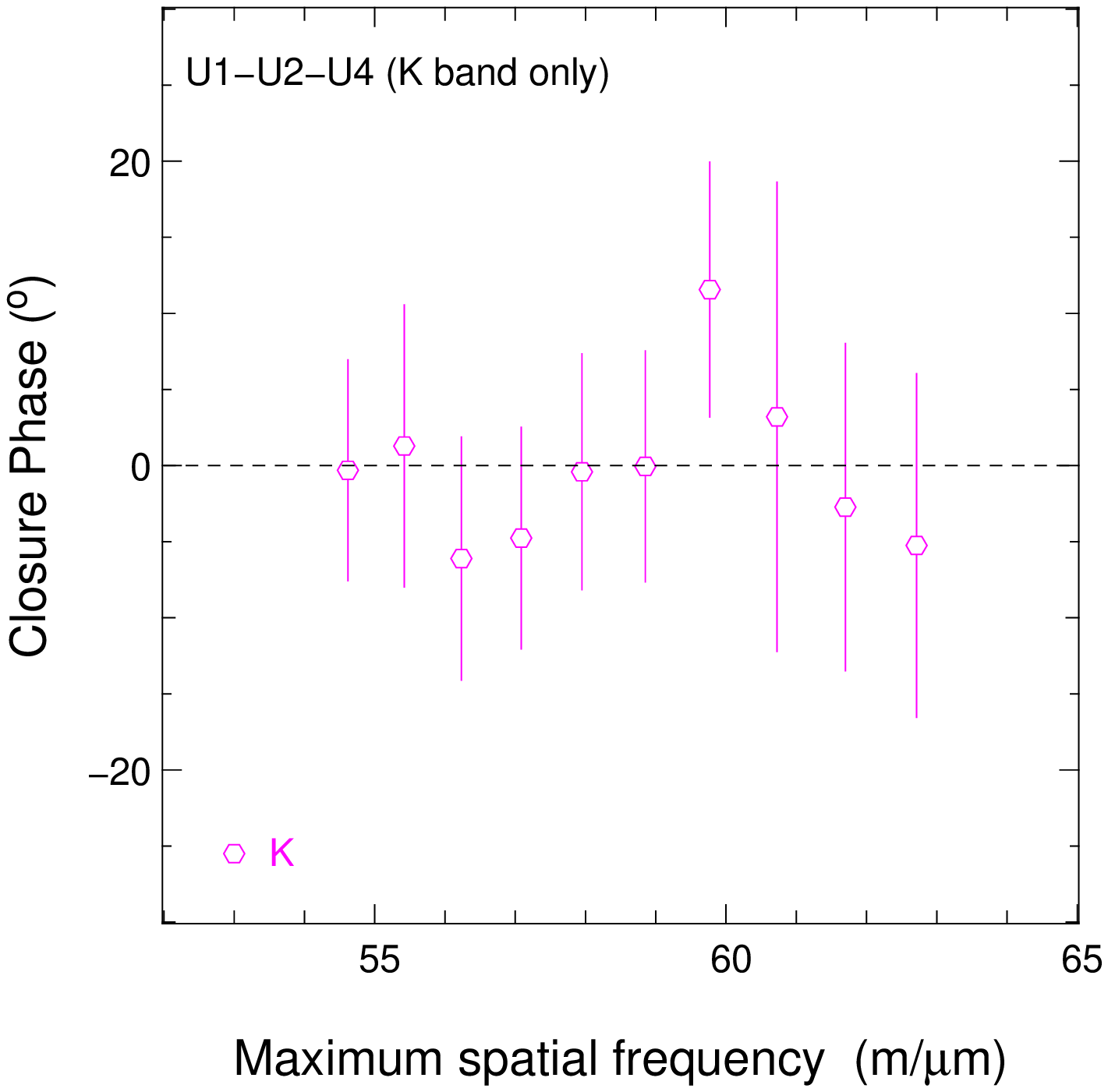}\\
\end{tabular}
\caption{\label{cpresoindiv} Measured closure phases for each
configuration  (reported in  the  top left  corners) plotted  versus
  maximum spatial  frequency.  In all panels, except  for U1-U2-U4 for
  which only $K$ band data were useful, the measurements include both
  the $H$ and $K$ bands.}  
\end{figure*}

\subsection{Closure phases}
The closure phase is a quantity that can be derived from 
interferometric observations with at least 3 telescopes. By combining up
the phases of the fringes  obtained with 3 telescopes, the atmospheric
disturbances  are cancelled out.  Consequently, the  sum of  the three
phases (the  closure phase) is atmosphere-free, \textit{i.e.}, independent of the
phase fluctuations.  It  is related to the degree  of asymmetry of the
observed brightness  distribution: a point-symmetric  object will have
zero closure phases, while a non-zero measurement is indicative of a deviation
from point-symmetry. The sign of the CP is derived from the way that it is
calculated (clockwise or counterclockwise).  Physically, the CP depends on 
the directions  that are  sampled by the  individual baselines  of the
configuration since they probe different asymmetries of the brightness
distribution, which in  turn strongly depend on the  ratio of the star
to circumstellar matter flux. 

One or more closure phase (CP) measurements were obtained for each
observing night, except for the night of 2008 May 26.  In total, we
obtained 304  CP in  the $K$ band,  and 128  in the $H$  band. This  is an
enormous improvement over the existing datasets for a young star. 
The full  set of  measured CP  across the $K$  and $H$  bands is  shown in
Fig.~\ref{cpresoindiv}, plotted against the maximum spatial frequency
B$_{\rm{max}}/\lambda$, where B$_{\rm{max}}$ refers to the projected length 
of the longest baseline in the corresponding configuration. 

\begin{figure*}[t]
\centering
\begin{tabular}{cc}
\includegraphics[width=0.4\textwidth]{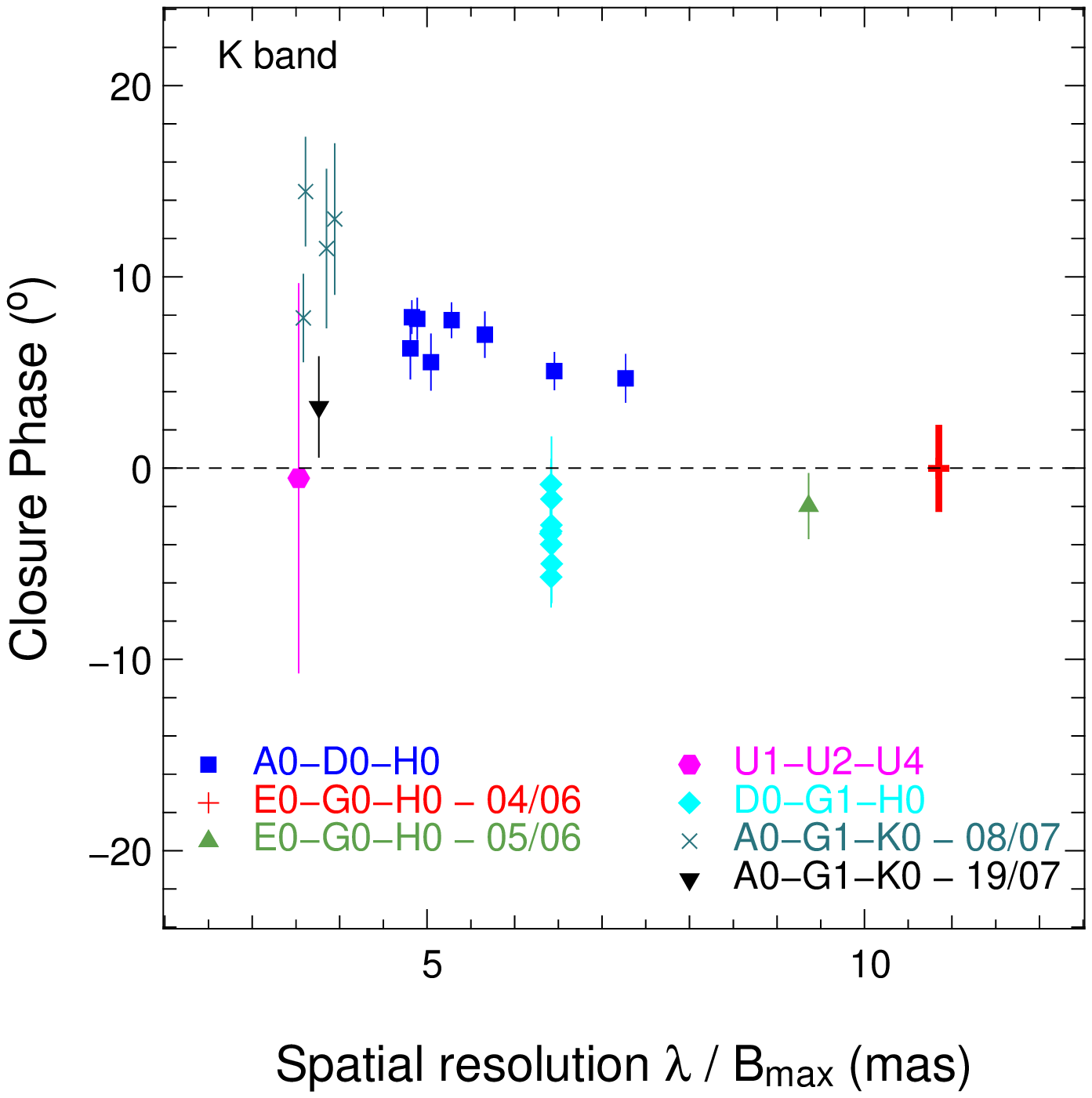} & 
\includegraphics[width=0.4\textwidth]{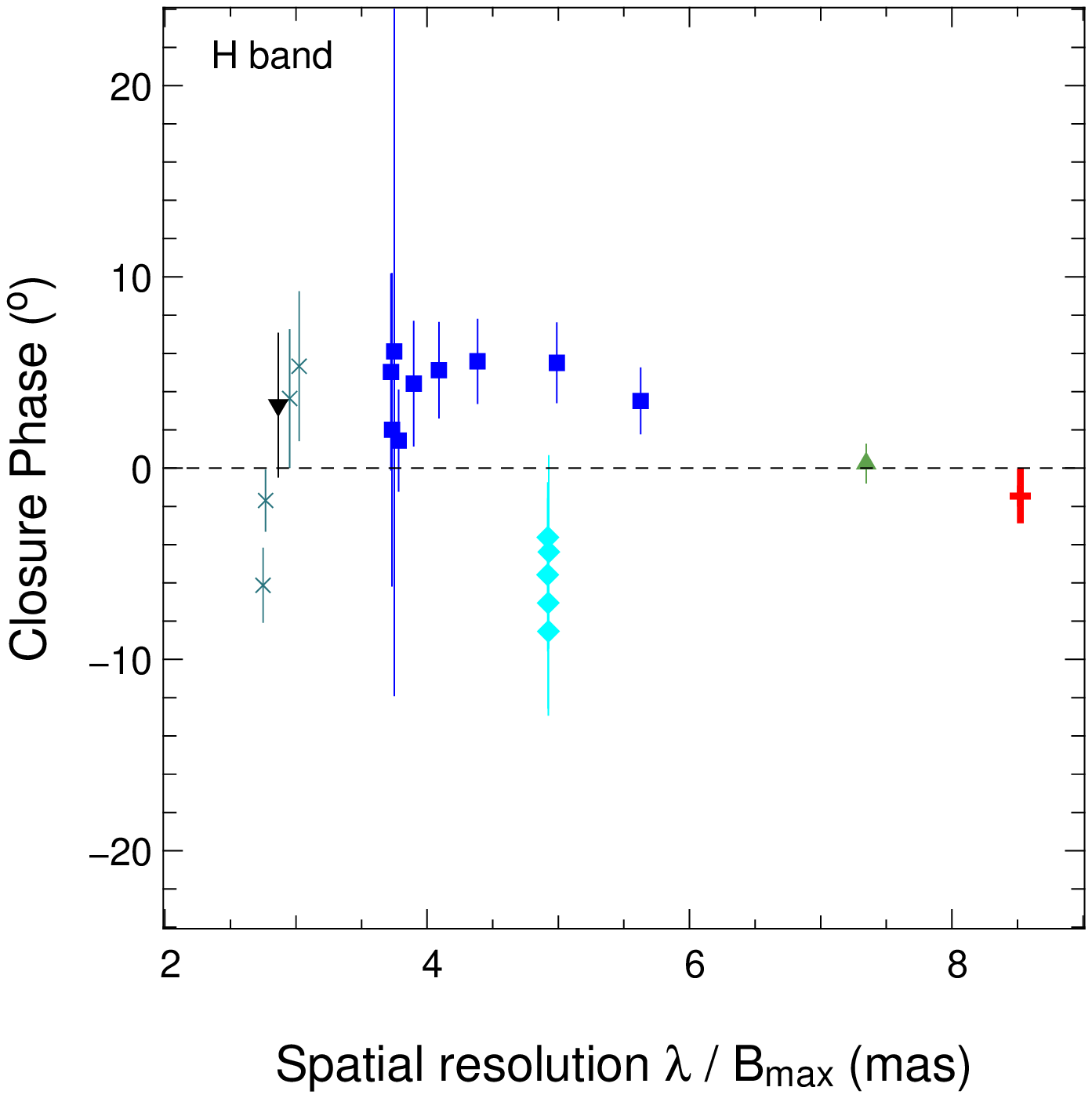}
\end{tabular}
\caption{\label{cp_reso_all} Broad-band closure phases plotted
  against the maximum spatial resolution achieved for all telescope configurations ($K$ band, left; $H$ band, right).  The
configurations  and the  corresponding symbols  are reported  at the
  bottom of the left panel.}  
\end{figure*}

The  CP signal  depends  on  the spatial  resolution  achieved by  the
interferometer    (unresolved     sources    are    centro-symmetric).
Figure~\ref{cp_reso_all}  show  the broad-band  CP  in  $K$ and  $H$
plotted      against      the      maximum     spatial      resolution
$\lambda/\rm{B}_{\rm{max}}$,    achieved   with    the   corresponding
three-telescope configuration.  It can be seen that the  level of the
CP signal increases with the power of resolution (see, 
for example, the configurations  of aligned telescopes along  the same
B$_{\rm{PA}}$, E0-G0-H0, and A0-D0-H0). 
Table~\ref{CP} gives the corresponding broad-band CP averaged over all
the measurements  as well as  the achieved maximum  spatial resolution
for each telescope configuration.  Since  we checked that all the CP were
calculated in the 
same  way,  the  relative  signs between  measurements  obtained  with
various configurations are relevant  and should be reproduced by model
fitting.  

The CP is  close to zero for the short linear  array (E0-G0-H0): we do
not detect any meaningful deviations from 
zero  (\textit{i.e.},    any  flux   asymmetries)  at  these   low  spatial
resolutions (10.8 and  8.5 mas in the $K$  and $H$ band, respectively).
\cite{monnier06} detected CP signals below 
$\leq 5^\circ$ for 12 Herbig~AeBe stars with the IOTA-3T
interferometer,   including   HD~163296   for   which   they   derived
0.6$\pm$0.4$^\circ$ at a resolution  of 11.8 mas.  Our observations at
similar resolution provide comparable results.  

At higher  resolutions, we  found that the  CP signal  is not
zero, \textit{i.e.}, that the emission is no longer centro-symmetric. However,
the departure of the closure phase  signals from zero is small, with a
maximum average CP in $K$ band of 11.7$\pm$2.9$^\circ$ and in $H$ band
of -5.8$\pm$2.5$^\circ$  (see Table~\ref{CP}). 
As we discuss in the following section, this level of asymmetry
is  not   compatible  with  strongly  skewed   distributions  for  the
circumstellar material in the innermost regions surrounding the star. 

Except in one case (A0-K0-G1), the  CP does not vary much with varying
hour angle  (see~Fig.~\ref{cpmod}).  This means  that, considering the
change  in maximum  resolution  and position  angle  that occurs  when
varying the hour angle (\textit{i.e.},  when changing the direction and
the projected  baseline length  B$_{\rm{L}}$), the level  of asymmetry
does  not change  much.  Consequently,  the circumstellar  matter must
have a rather smooth azimuthal brightness distribution.

The $H$ band CP are slightly lower than those measured in the $K$ band,
as expected since the emission in $H$ probably originates in a more
compact region, as most of our data suggest, than the one emitting the $K$ band
flux.  However, considering  the  large error  bars,  this effect  is
hardly significant.  Over  the $K$ band and $H$  band separately, 12\%
and 15\% of the CP, respectively, show a variation above the 
1-$\sigma$ level and all measurements are consistent with variations 
within the 2-$\sigma$ level. Considering the entire range of wavelength
($H$  and  $K$  together),  45\%  of  the CP  show  a  variation  with
wavelength above the 2-$\sigma$ level.

\begin{table}[t]
\centering
\caption{\label{CP} Values  of the closure phase as  averaged over all
  the  measurements  obtained with  the  various configurations.   The
  dates of the observations, the wavelength range, and the
achieved  maximum spatial resolution  are reported. }
\begin{tabular}{c c c c c}
\hline
\hline
Configuration & Date & Band & Resolution & CP \\
 & & & [mas] & ($^\circ$) \\
\hline
A0D0H0 & 24/05/08 & $K$ & 7.9 & 6.5$\pm$1.4 \\
 & & $H$ & 5.8 & 4.3$\pm$1.8 \\
E0G0H0 & 04/06/08 & $K$ & 11.6& 0.0$\pm$2.1 \\
& & $H$ & 9.0 & -1.5$\pm$1.3 \\
& 05/06/08 & $K$ & 10.0& -2.0$\pm$1.7 \\
& & $H$ & 7.9& 0.2$\pm$1.1 \\
U1U2U4 & 24/06/08 & $K$ & 3.8 & -0.5$\pm$10.1 \\
D0H0G1 & 06/07/08 & $K$ & 7.0 & -3.3$\pm$1.6 \\
& & $H$ & 5.1 & -5.8$\pm$2.5 \\
A0K0G1 & 08/07/08 & $K$ & 3.3& 11.7$\pm$2.9 \\
& & $H$ & 3.2 & 0.3$\pm$4.6 \\
& 19/07/08 & $K$ & 4.1& 3.2$\pm$2.6 \\
 & & $H$ & 3.0 & 3.3$\pm$3.7 \\
\hline
\hline
\end{tabular}
\end{table}

\section{Modelling the interferometry results }
The analysis of visibilities and closure phases requires the assumption
of a model for the brightness  distribution on the plane of the sky to
be  compared  with  the  observations.  Even when  the  comparison  is
successful, it  is impossible to know  if the solution  is unique, and
because models are difficult to compute, it is practically impossible to
explore  all  possibilities.  Here,  we  are guided  by  the  current
paradigm that  the NIR emission in  excess of the  photospheric one is
produced by the inner parts of a circumstellar disk.  

\subsection{Stellar parameters}
\label{sec:stellar}
To model the interferometry results, one needs to know
  the unresolved  contribution of  the star to  the total flux  at the
  wavelength of  interest.  We compute it from  the observed magnitude
  in  the I  band (\textit{i.e.},  6.71 $\pm$  0.026, with  very little
  variability  over  a  period  of  about  20  years  \citep{deWinter01,
  tannirkulam}, assuming ZAMS colors for  a A1 star and an extinction
  $A_V=0.25$.  The resulting stellar fluxes  are about 1.4 Jy in K and
  2.2  Jy  in H.   The  observed  NIR  fluxes display  a  moderate
  variability \citep{sitko08}, so  that the stellar contribution may
  be in  the interval 14--18\% in  K and 33--37\%  in H, respectively.
  The  effect of  variability on  the interferometry  results  and the
  desirability of performing  simultaneous photometry were discussed
  by \cite{sitko08}.  This, unfortunately, is practically impossible
  with AMBER/VLTI, and we do not  know the values of the total flux at
  the  time of the  observations.  In  this paper,  we adopt  a stellar
  contribution of $\sim$14\%  and $\sim$33\% to the observed $K$ and $H$
  band fluxes,  respectively, \textit{i.e.}, on  the lower side  of the
  estimates. These numbers are in agreement with the CHARA and AMBER 
measurements  at long  baselines, where  all the  circumstellar matter
  appears to be resolved, and are similar to the values adopted by
\cite{tannirkulam}. The HD 163296 SED is shown in Figs.~\ref{gas} and~\ref{dust}.

\begin{figure*}[t]
\centering
\begin{tabular}{cc}
\includegraphics[width=0.45\textwidth]{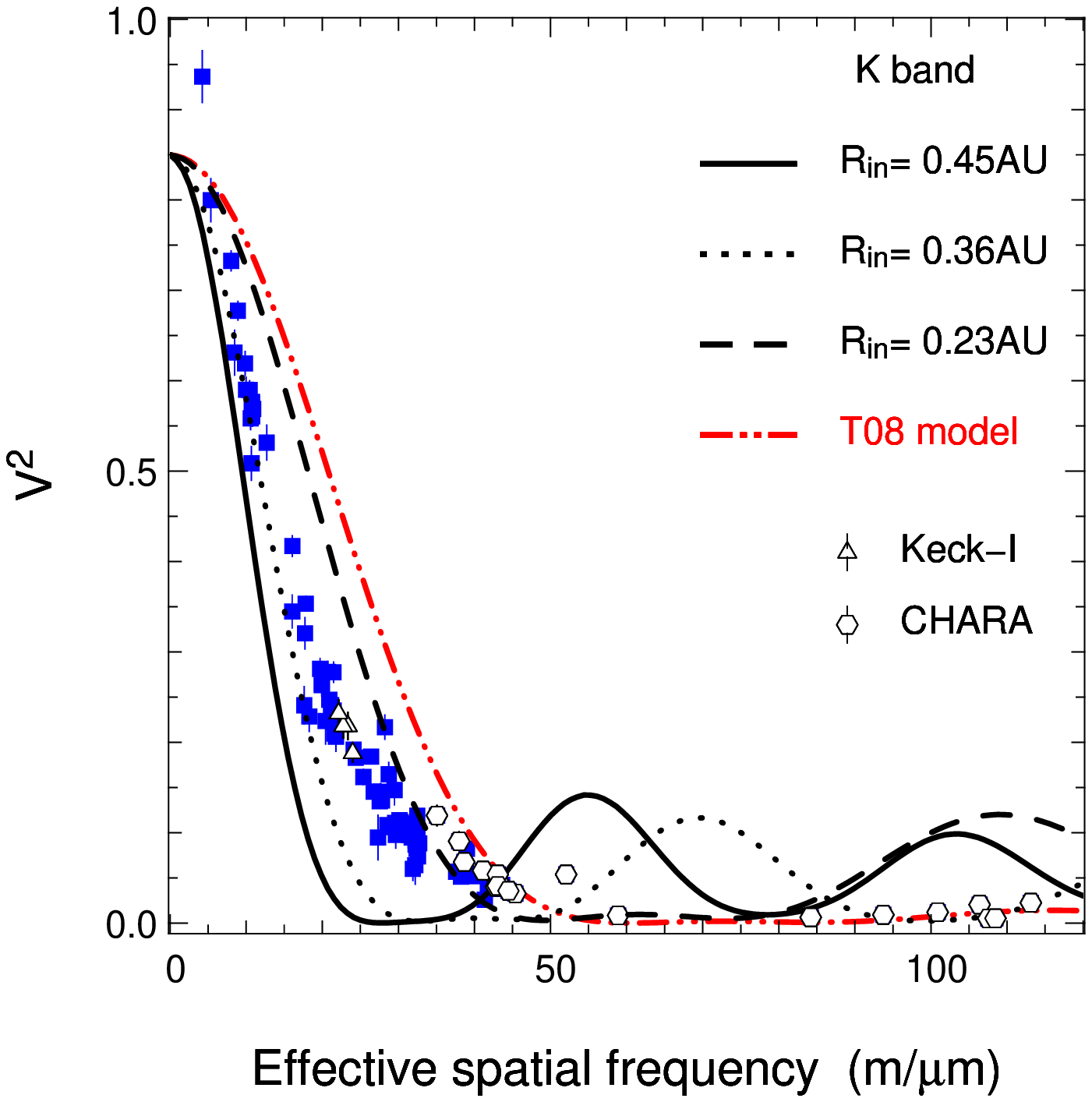} & 
\includegraphics[width=0.44\textwidth]{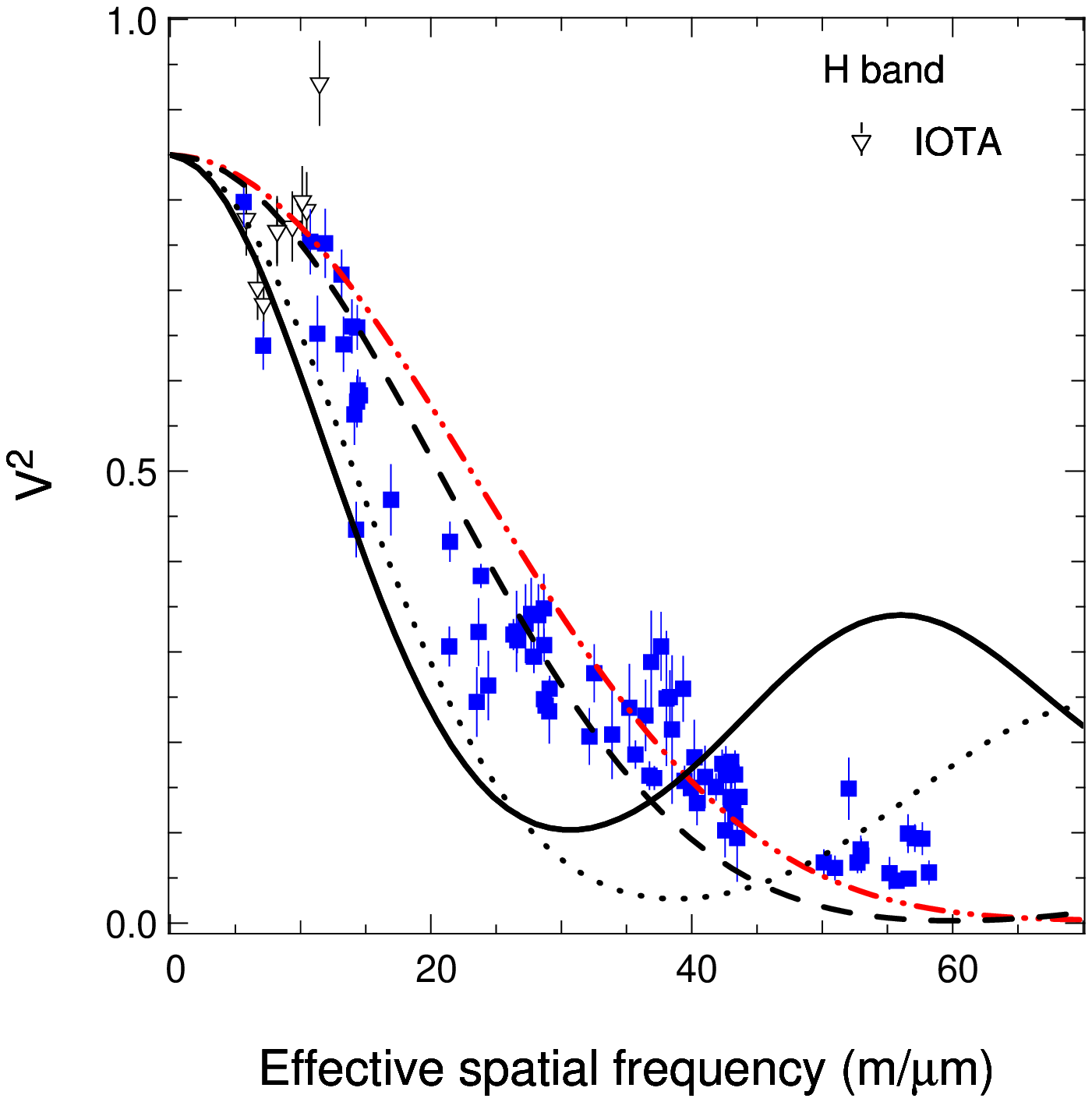}\\
\end{tabular}
\caption{\label{fig:rimonly} 
Model visibilities when considering a rim alone, located
at  0.23, 0.36, and  0.45~AU (dashed  red, dotted  blue, and  full black
lines respectively), versus effective spatial frequency B$_{\rm{eff}}/\lambda$, compared to the measured
broad-band visibilities (blue full squares) in the $K$ band (left) and
the $H$ band (right).  The predictions of the model described in
T08, which  includes an additional inner component  interpreted as gas,
is  given  in red  (T08;  dash-dot-dot).   The  $K$ band  measurements
obtained  at  CHARA and  Keck-I  are  overplotted  (circles  and
triangles respectively) in the left panel, as well as the IOTA $H$
band data  (triangles) in  the right panel.  An  incoherent flux
contributing  to $\sim$8\% of  the total  $H$ and  $K$ band  fluxes is
considered. }  
\end{figure*}

\subsection{Incoherent flux}
In our models,  we consider incoherent  flux possibly
emitted by  an extended halo,  as first suggested  by \cite{monnier06}. 
They estimated its contribution to be 5\% of the total
$H$~band  fluxes (\textit{i.e.},  10\%  in V$^{2}$),  a value  that T08
also used when modelling their $K$~band data. 
Our  $H$~band  data  suggest  a  slightly higher  value  of  $\sim$8\%
(\textit{i.e.},   $\sim$15\% in  V$^{2}$).   We adopt  this value  when
fitting both  the $H$  and $K$~band interferometric  observations. The
precise origin of this emission is unknown and discussing it is beyond
the scope of this paper.  

\subsection {Disk position angle and inclination}
Values of  inclination (i)  and position angle  (PA) of the  HD 163296
disk (\textit{i.e.},  its major axis)  have been derived  at different
wavelengths with a variety of techniques and tend to agree: 
\cite{wassell06}
(i=51$^\circ$$^{+11^\circ}_{-9^\circ}$; PA=139$^\circ\pm$15$^\circ$),
\cite{isella07}  (i=46$^\circ\pm$4$^\circ$;PA=128$^\circ\pm$4$^\circ$)  and
T08 (i=48$^\circ\pm$2$^\circ$;PA=136$^\circ\pm$2$^\circ$).  
They  are also  consistent with  our visibility  data, which  infer an
inclination of $40^\circ \pm 10^\circ$ and a position angle of 
$140^\circ \pm  15^\circ$, when fitted  with geometrical models  of 
uniform brightness  ring.  Since we do not  constrain these parameters
better, in the following, we adopt the 
values    derived   by    T08,    \textit{i.e.},    i=48$^\circ$    and
PA=136$^\circ$. In the following sections, we use the effective
baseline, which is defined by~: 
$$B_{\rm{eff}}=                B\sqrt{cos^{2}(\theta)                +
  cos^{2}(\Phi)sin^{2}(\theta)}$$ 
where $\theta$ is the angle between the baseline direction and the major
axis of  the disk, and $\Phi$  is the disk  inclination (T08).  This
representation allows us to show  the data in  a concise way  once the
inclination and position angle of the disk are known.

\subsection{A disk rim}
Early NIR interferometric  studies of  Herbig~Ae stars  have shown
that standard  accretion disks, extending  up to the  dust sublimation
radius do  not fit the  observations \citep{ppv} and that  superior fits
are obtained by assuming that the disk develops a curved rim, probably (but not
necessarily) controlled by dust  sublimation and its dependence on gas
density  \citep{isella05,tannirkulam07}.    Not  only  the   disk  rim
hypothesis      is     supported     by      physical     calculations
\citep{dullemond01,isella05}: the overall properties of HD~163296 
are also consistent with this  model. The SED appears to be consistent
with a rim emission up to 7-8~$\mu$m and  a disk in the shade at
longer wavelengths \citep{dullemond01,isella07}. In addition, our measured visibilities show a
dependence on hour angle  (or similarly on baseline position angle
B$_{\rm{PA}}$)  consistent with  an elliptical  shape, as  expected in
either a rim or any axisymmetrical distribution seen at high
inclination.  Finally,  the   detection  of  non-zero  closure  phases
supports the idea that an asymmetric brightness distribution - such as
the disk  inner rim  - contributes to  the NIR
emission.  \\
As in the NIR, the contribution of the disk outside the rim  
can be neglected \citep{isella07}, we begin by examining a star 
+ rim system to model the NIR emission following \cite{isella05}. 
We  adopt a stellar  luminosity and  mass of  30~L$_\odot$ and
  2.3~M$_\odot$,  respectively,   and  an  effective   temperature  of
  9250~K. The  disk is assumed  to be in hydrostatic  equilibrium. The
  dust consists of silicates with optical properties given by
\cite{weingartner01}.  The evaporation temperature, of around 1500~K,
  depends on the local gas density as in \cite{pollack94}.  Since the 
  shape of the rim is controlled  by the largest grains, we consider a
  single size for the silicate dust, which is therefore the only free
  parameter  in   the  model.   The  dependence   of  the  evaporation
  temperature on $z$ implies that  the distance from the star at which
  dust evaporates  increases with $z$,  \textit{i.e.},  that the  rim is
  curved.  

As    can  clearly   be   seen   in  Fig.~\ref{fig:rimonly},   our
interferometric measurements are inconsistent with a circumstellar
emission produced by a rim only, regardless of its location.  This figure
shows the AMBER broad-band visibilities as a function of the effective 
spatial frequency  B$_{\rm{eff}}/\lambda$.  In addition to the
AMBER  data, we  include  the  $K$ band  visibilities  obtained at  CHARA
\citep[circles;][]{tannirkulam} and  at Keck-I \citep[triangles;][]{monnier05} as  well as the
$H$ band IOTA data \citep[upside down triangles;][]{monnier06}.  We emphasize that all measurements from
these four interferometers are compatible with each other within their
error bars.  The results for three rim models computed by assuming a 30~L$_{\odot}$  stellar luminosity and silicate grain sizes of 3,
0.6, or  0.3~$\mu$m, corresponding  to inner rim  radii at
0.23, 0.36,  and 0.45~AU, respectively, are  overplotted.  These models
produce  an emission that  does not  exceed the  observed flux  at any
wavelength, and, contributes respectively, to about 80, 50, and 28\% of
the observed emission in $H$, and 91, 86, and 66\% in $K$.  All models
show large  departures from the observations.  In  addition, all these
models have very asymmetric emission, and produce a closure 
phase signal  greater than  observed (see Fig.~\ref{cpmod},  red dashed
lines).  Our large data sets, which include $H$ and $K$ visibilities
and closure phases, reinforce the conclusions of other authors, namely
that there should  be an additional, symmetric emission  closer to the
star (T08).  

\subsection{A bright inner disk}
\begin{figure*}[t]
  \centering
  \begin{tabular}{cc}
    \includegraphics[width=0.44\textwidth]{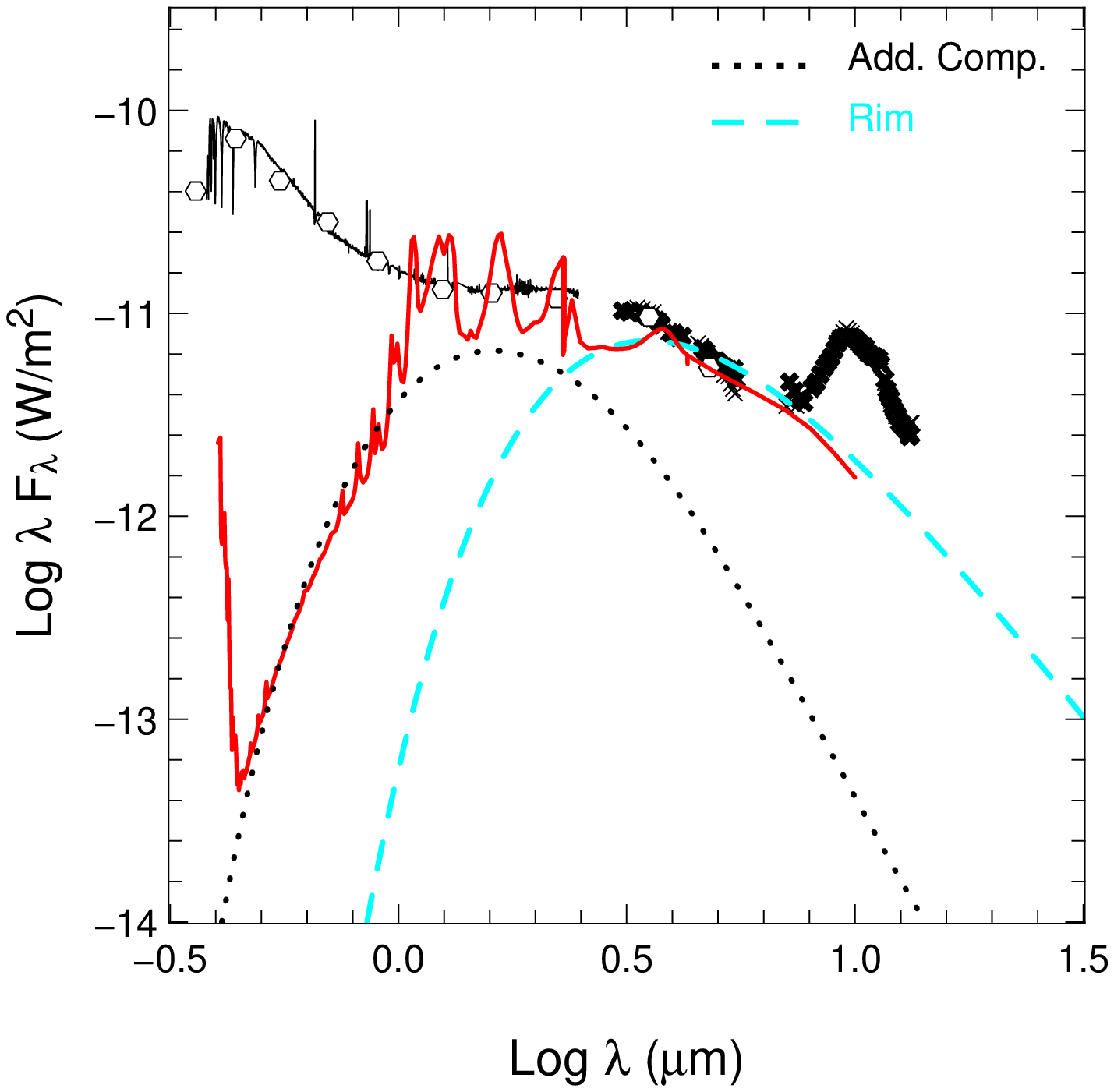}
    & 
    \includegraphics[width=0.44\textwidth]{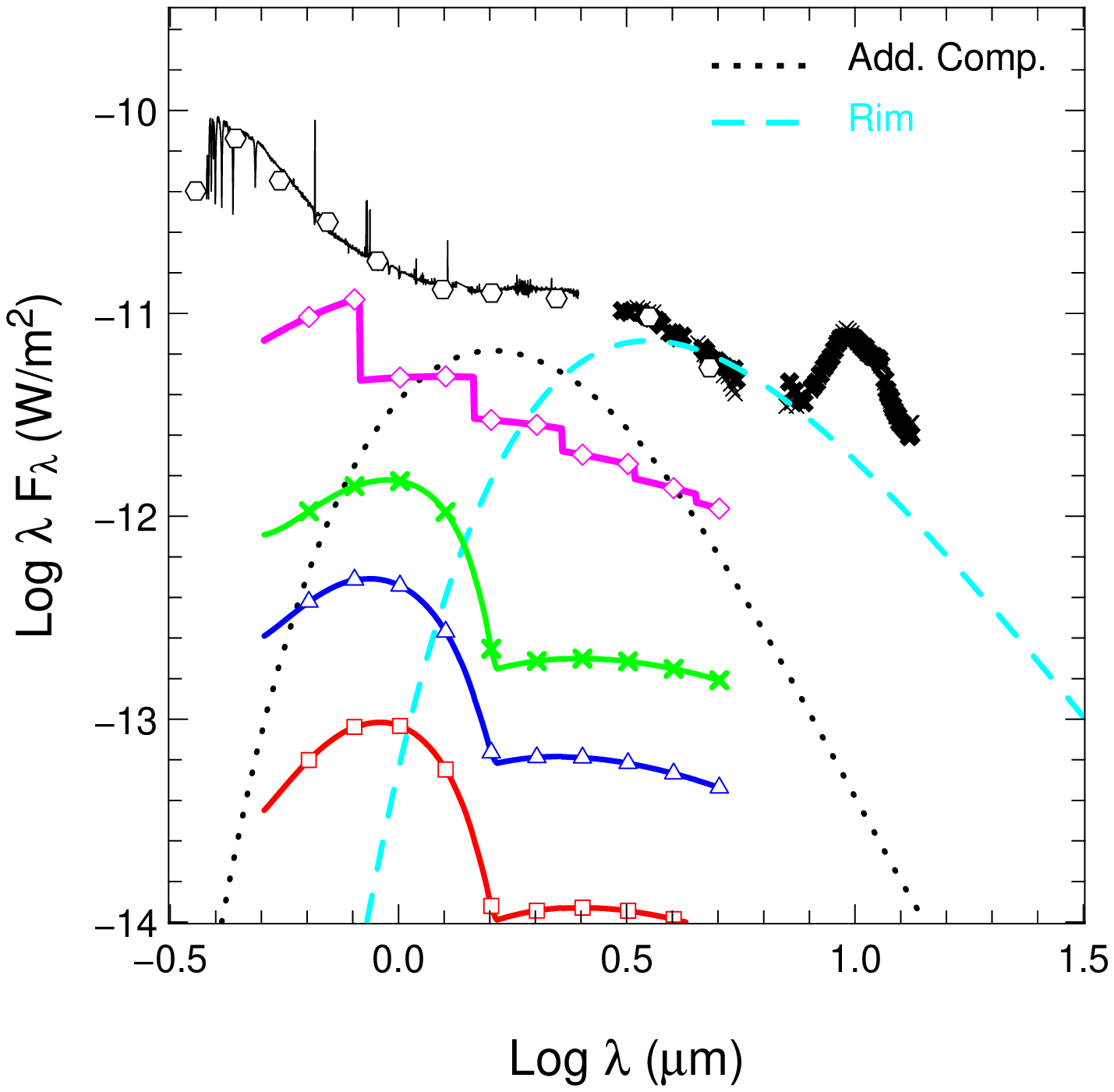}\\
  \end{tabular}
  \caption{\label{gas} SED of HD~163296 \citep{sitko08, tannirkulam}
    with the predictions of gaseous disk models. The flux emitted by 
    the rim located at 0.45~AU (cyan dashed line) as well as the additional
    inner disk emission (black dotted line) are added in both panels to
    allow  direct  comparisons.   Left:  the  predictions  for
    optically thick and dense layers of gas in LTE \citep{muzerolle04}
    are overplotted (red full line). In this case, the gas is mostly in
    a molecular state. 
  Right:  the continuum  emission  as predicted  by thin
    disks of gas in non-LTE, heated from the top by the
stellar radiation, are shown. The gas extends
from 0.1 to 0.45 AU and has constant surface densities of 0.1, 1, 
    and 6  g/cm$^2$ (red line  with squares, blue line  with triangles
    and green line with  diamonds, respectively).  In these conditions,
    the gas is mostly atomic. The predictions of a fully ionized layer
    of  gas  at  8000~K  with  a constant  surface  density  of  0.06
    g/cm$^{2}$  are also  reported (magenta  curve with  diamonds). As
    long  as  the  disk   remains  optically  thin,  the  emission  is
    proportional  to the surface  density, and  this value  was chosen
    only to display its wavelength dependence.} 
\end{figure*}

A rim   alone cannot reproduce our observations.   As a
  matter of fact, the lack  of bounces in the visibilities suggests
  a continuous and smooth distribution  of matter in the inner regions
  of HD~163296, and a low contribution of the rim to the NIR emission.
  In the following, we explore the effect of adding emission
  from  an  inner   disk  (inside  the  rim)  to both  the  star  and  rim
  contributions.  We fix the rim location to be R$_{\rm{rim}}$=0.45~AU. 
  This   large  radius   is  implied  by   the  results   of  our
  three-component models, as  we show in the  following. It can,
  for example, be obtained by considering small silicate 
grains (from  0.05 to 0.5~$\mu$m)  in a high-density dusty  disk, or
  micron-sized grains  in a low-density region,  where the evaporation
  temperature  is  lower.   In  both cases,  the  rim  effective
  temperature is quite low (T$_{\rm{eff}}\sim$1000~K) and accounts for
  the total observed flux in  the wavelength range 4-8~$\mu$m, but not
  at shorter wavelengths (see  Fig.~\ref{gas}, cyan dashed line).  The
  SED  of   the  proposed  additional  inner  disk   can  be  computed
  by subtracting both the stellar and  disk rim fluxes from the observed
ones.  As shown in Fig.~\ref{gas} (dotted line), it peaks in the $H$
  band and is higher than  the rim emission at all wavelengths shorter
  than about 2.5~$\mu$m.  Therefore,  the NIR emission is dominated by
  the inner disk component rather than the rim. \\ 
  We obtain a first estimate of the emission properties of the inner disk
   by modeling it as a region of constant surface brightness
between an inner and  outer radius.  The surface brightness is 
  constrained by the condition that  the integrated flux must be equal
  to the flux  derived from the SED for  the additional component.  We
  compute  visibilities  and find  that  these three-component  models
  reproduce  the $H$  and $K$  band observations  quite well  over the
  entire  range of  baselines  if the  smooth,  inner emission  extends
  between $\sim$0.10  and $\sim$0.45~AU.  The emission  can be roughly
  described  as that of  a diluted  black-body with  temperature $\sim
  1600$ K and optical depth $\sim$0.2, which decreases as
$\lambda^{-1.6}$    with    increasing    wavelength.    A    natural
  interpretation is that the  emission originates in an optically thin
  region inside the  rim.  In the following sections,  we discuss
  the physical nature of this component.

\subsubsection{An inner gaseous disk ?} 
Several authors have suggested that NIR interferometers detect the
emission    of    gas    within    the   dust    evaporation    radius
\citep{eisner07,isella08,tannirkulam,kraus08}.    This  interpretation
presents  several difficulties  in the  case of  HD 163296,  where the
additional component dominates the emission  in $H$ and $K$, and where
good observations of high spectral resolution exist over a wide range
of wavelengths.  

Models of the emission of purely gaseous disks inside the dust
sublimation radius  were  computed by \cite{muzerolle04}  for HAe
stars assuming LTE opacities.  For typical HAe accretion rates ($\sim$
$10^{-7}$~M$_{\odot}$yr$^{-1}$), the inner, dust-free, disk gas 
surface density  is higher than  $10^3$ g/cm$^2$, the  gas temperature
ranges from a few thousand to a few hundred K, and the gas is fully
molecular.  The NIR emission is sufficiently strong to, in
principle, account for the observations. However, as has been pointed out by
several authors, the  models also predict many strong molecular
bands (mostly water  and CO overtone transitions) that  are absent in
the     HD    163296     spectrum    (see     Fig.~\ref{gas},    left)
\citep{najita00,najita07}.    This  problem   is  also   discussed  in
\cite{najita09}, in  connection with the non-detection  of CO overtone
and water emission in the Herbig~Ae star, MWC480, that also exhibits a hot
compact  NIR  excess   previously  detected  with  interferometry  and
interpreted as resulting from water \citep{eisner07b}.  

The assumption that the gas is in LTE, however, is certainly inappropriate
at least in the upper disk layers, where the stellar radiation
can penetrate, ionize, and dissociate matter well above the LTE predictions.
However, these thin gas layers are unlikely to contribute significantly to the
broadband  observed fluxes. 
We  computed  the  emission  from  thin  layers  of gas,  
using the code Cloudy \citep{ferland98}.  We assumed that the region
inside the rim  can be described as a  geometrically thin disk, heated
from the outside by the star. The disk extends 
from 0.1 to 0.45~AU and has a constant surface density. 
In these conditions, as long as the disk is optically thin to the
stellar  radiation,  the  gas  is  mostly atomic  and  the  ionization
fraction is low (\textit{e.g.},  $<0.01$). In the NIR, H$^{-}$ dominates
the  emission. We  show in  Fig.~\ref{gas}, right,
that for surface density values of 0.1, 1, and 6 
g/cm$^{2}$  (all  much lower  than  predicted  by the  viscous
  accreting disks  modelled by Muzerolle et al.  2004), the continuum
emission is always too weak  to be significant.  Increasing it further
would require  higher surface densities, in which  case LTE conditions
would very likely be reached.  In fact, for $\Sigma=6$ g/cm$^2$, the mean
optical depth to the stellar radiation is already of order unity.  
Moreover,  we  note that  the  wavelength  dependence  of the  non-LTE
continuum is inconsistent with the observations.  
In particular,  the H-  bound-free emission produces  a sharp  drop at
1.6~$\mu$m corresponding  to its activation energy, which  would be
seen in the HD~163296  high resolution spectra \citep{sitko08} if the
gas emission was higher.  

T08 suggested that hot gas is the physical origin of the additional
  inner  component. We  note that  their models  cannot  reproduce our
  AMBER/VLTI   data,  mostly   because  of   their  small   inner  rim
  radius. However, we examined the possibility that the gas is heated
  to higher  temperatures by  additional energy sources.   We computed
  the  properties  of  the  same  disk  model  but  after  fixing  the
  temperature to 8000~K. In this case, the ionization fraction is high ($\gg$0.5) and
  the emission is dominated by bound-free processes.  The emission can
  be very high,  but its wavelength dependence is  inconsistent with
  the observations (see Fig.~\ref{gas}, right).

 Based on these crude  considerations, we tend to exclude that
  the NIR  flux detected  by the interferometers  is dominated  by the
  emission of hot gas inside the dust sublimation radius.  However, it is
  clear that,  before this  can be definitely  ruled out,  one needs
  more realistic,  non-LTE models that treat  the transition from
  optically thin to optically thick layers, \textit{i.e.}, from atomic
  to molecular gas in a  dust-free environment.  These models would also
  be  important for  the  interpretation of  the  hydrogen and  helium
  recombination lines, which appear very strong in the models.  Atomic
  lines - mostly  hydrogen and helium ones -  are often interpreted as
  being  emitted in  magnetospheric accretion  columns of  gas.  This,
  however, is unlikely  to be true for most  Herbig~Ae stars, based on
  the   results  obtained   with  spectro-interferometry   around  the
  Br$_\gamma$  emission line  \citep{kraus08b,eisner09}, since this line
  seems to  be formed in  most cases, in  disk material closer  to the
  star than  the silicate dust sublimation  radius but outside
  the corotation radius.

 \begin{figure*}[t]
\centering
\begin{tabular}{cc}
\includegraphics[width=0.435\textwidth]{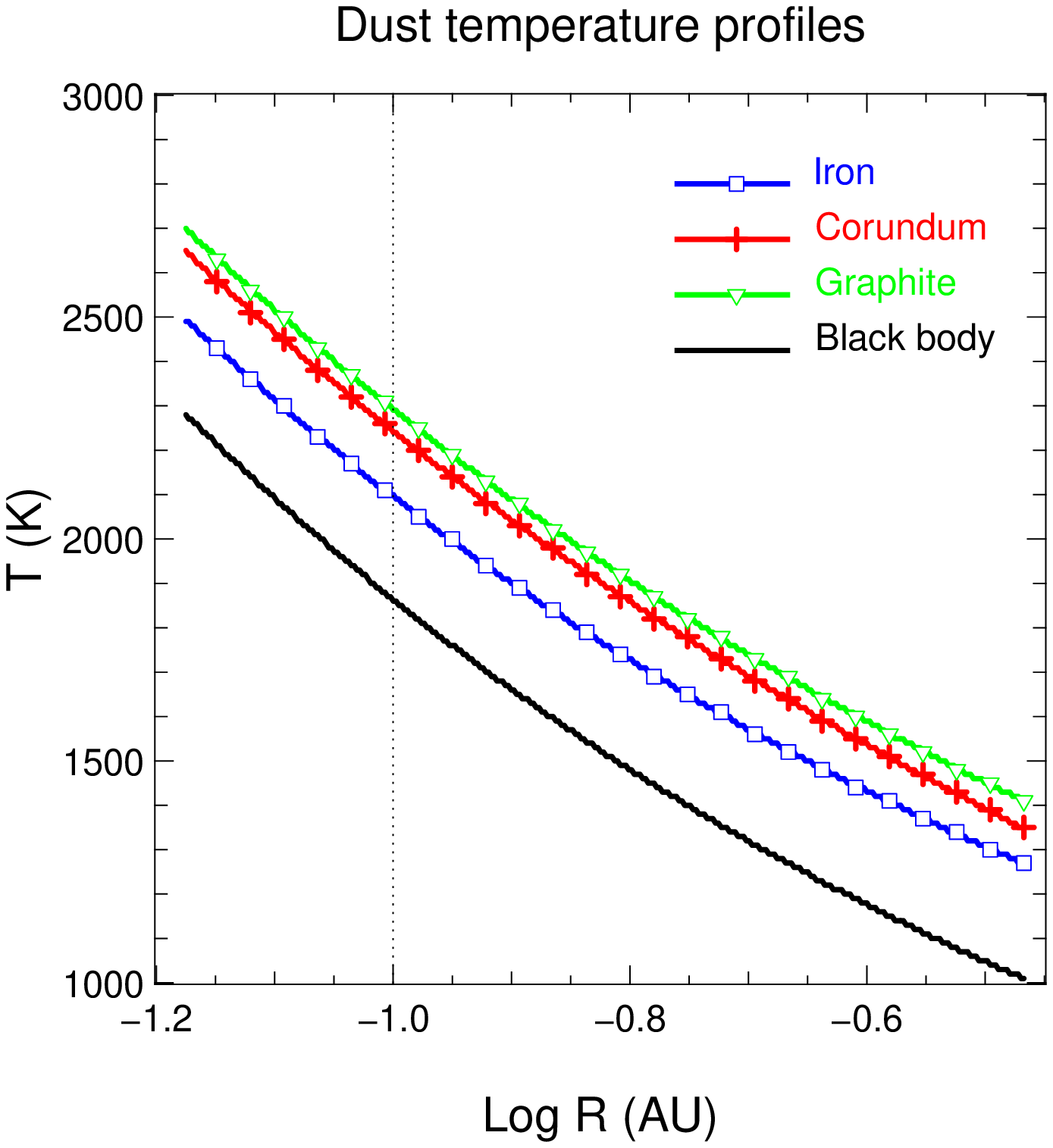}
& 
\includegraphics[width=0.44\textwidth]{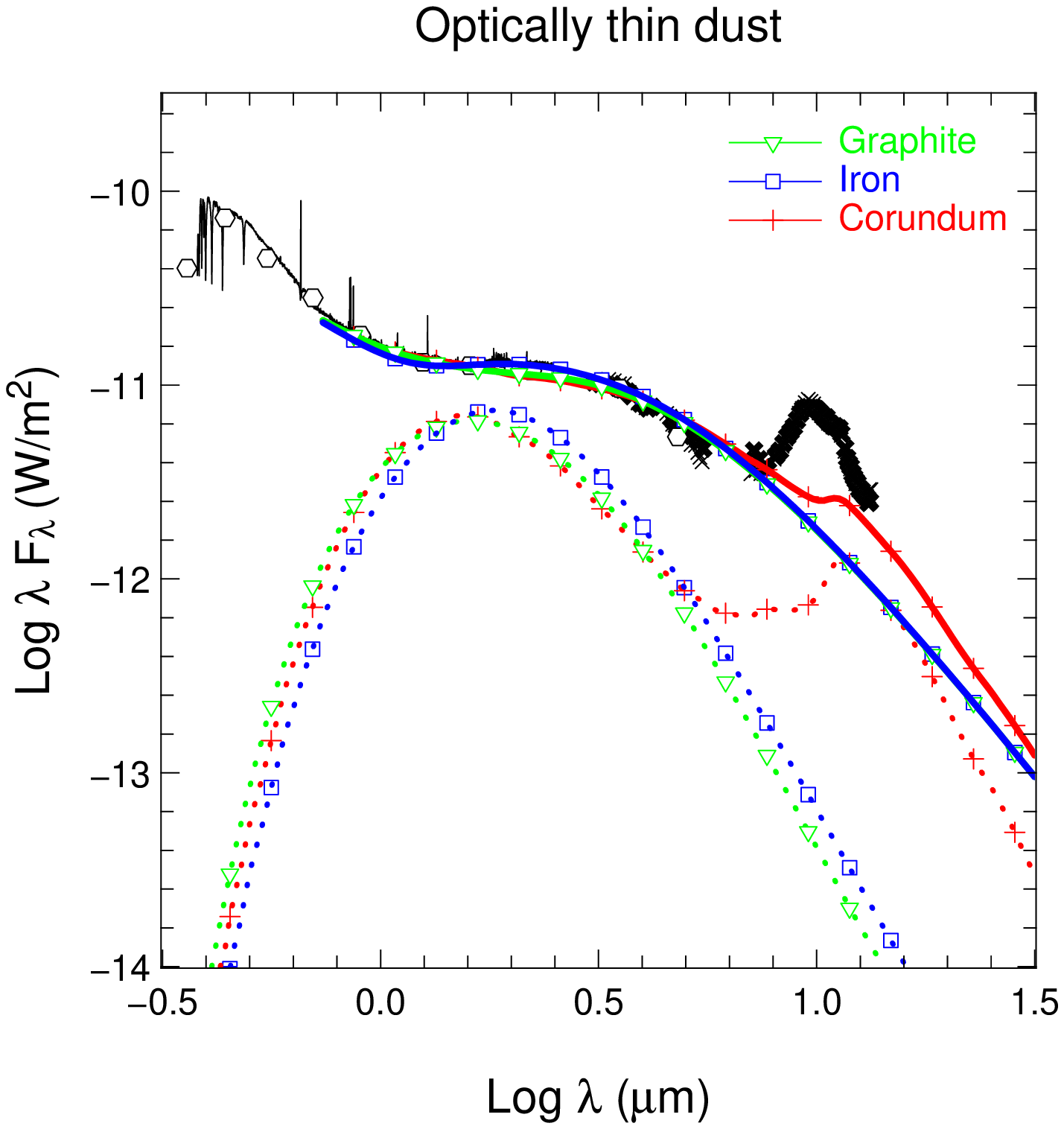}\\
\end{tabular}
\caption{\label{dust} Left:  radial temperature profiles  for graphite
  (green line with triangles), corundum (red line with
  crosses), or  iron (blue line  with squares).  The prediction  from a
  black  body (\textit{i.e.},   large grains)  is overplotted  to allow
  comparisons.  The vertical dotted line indicates 0.10~AU.  
  Right: SED of HD~163296 together with the predictions
  of three models  that include an inner disk made of
  dust in addition to the star and the rim.  The additional disk component has
  structural characteristics that are reported in Table~\ref{tab:bestmodels} and is
  made of  a single refractory  species - either graphite,  corundum, or
  iron.  The dotted lines  represent the  corresponding emission,  while the
  full lines show the total flux.}
\end{figure*}

\begin{figure*}[t]
\centering
\begin{tabular}{cc}
\includegraphics[width=0.445\textwidth]{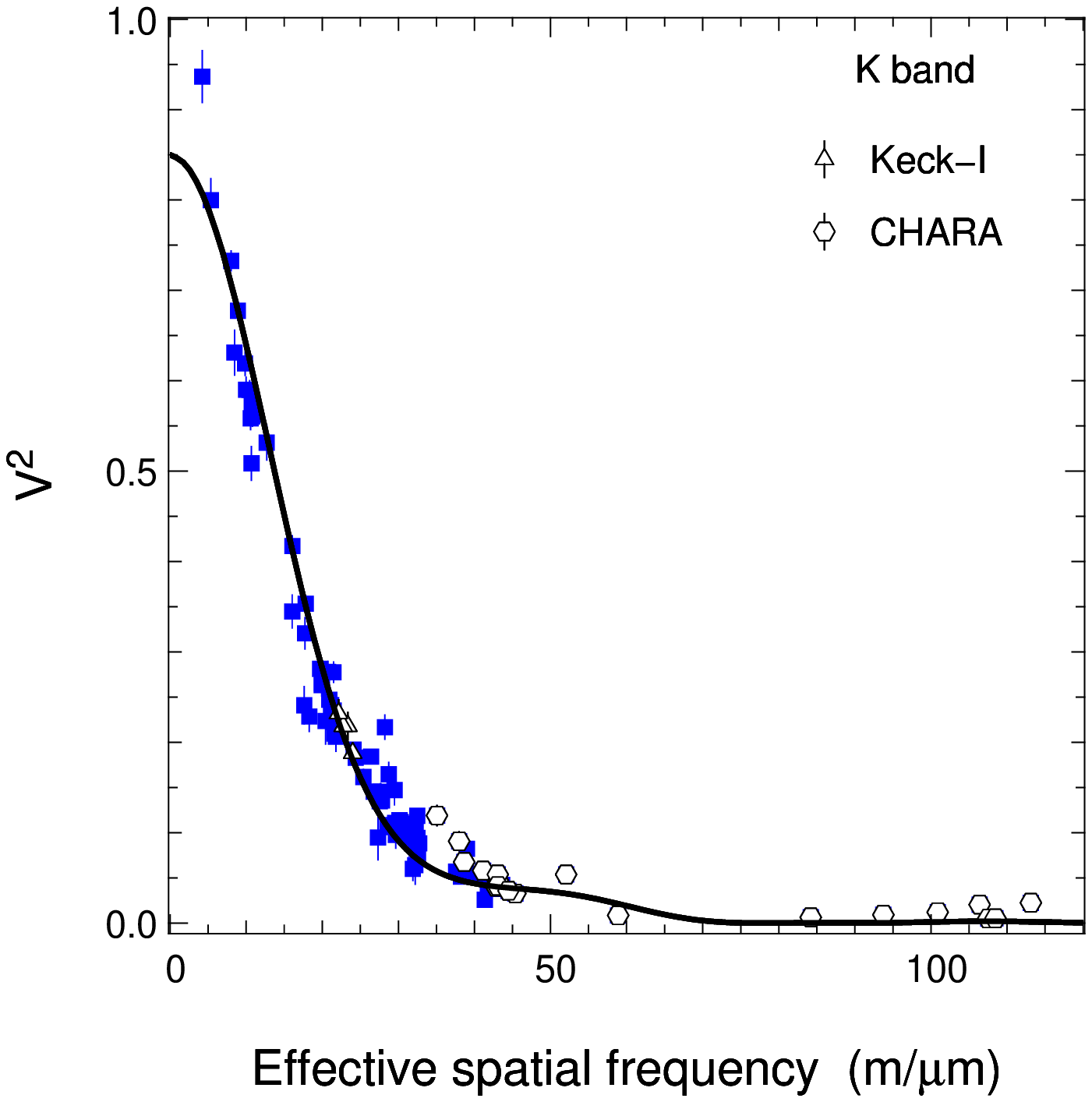}
& 
\includegraphics[width=0.44\textwidth]{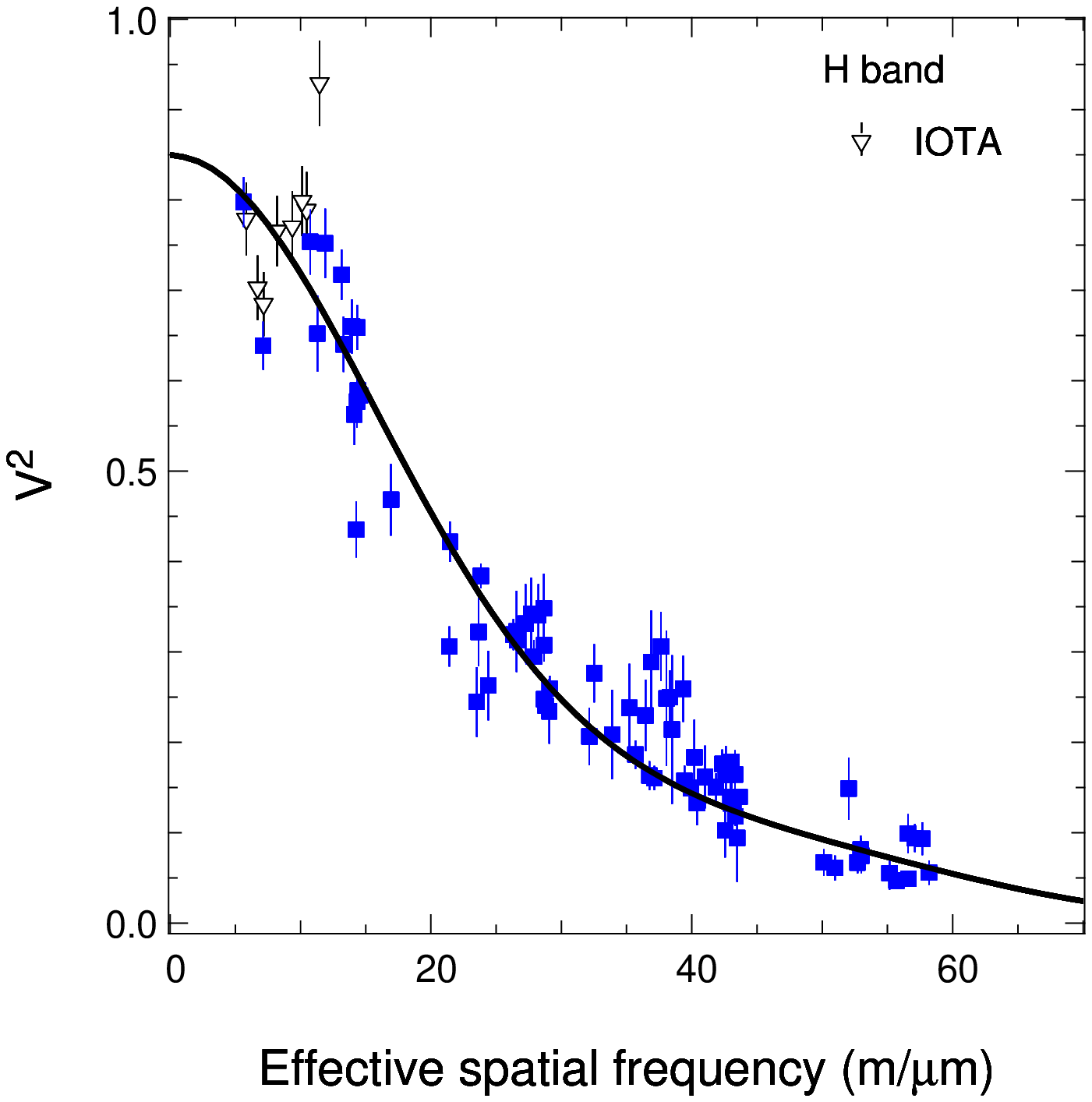}\\
\end{tabular}
\caption{\label{fig:bf} 
 Left:  Visibilities produced  by  our model  that  assumes that  iron
 grains are responsible for  the additional dusty inner disk emission,
 in $K$ band (left) and $H$ band (right) compared to the observations. 
 The AMBER/VLTI observations (blue
 full squares) as well as the Keck-I (triangles) and the
 CHARA (circles) are added.  The IOTA $H$ band data (upside down triangles) are
 plotted together with the AMBER visibilities. } 
\end{figure*}

\begin{figure*}[t]
\centering
\begin{tabular}{cc}
\includegraphics[width=0.42\textwidth]{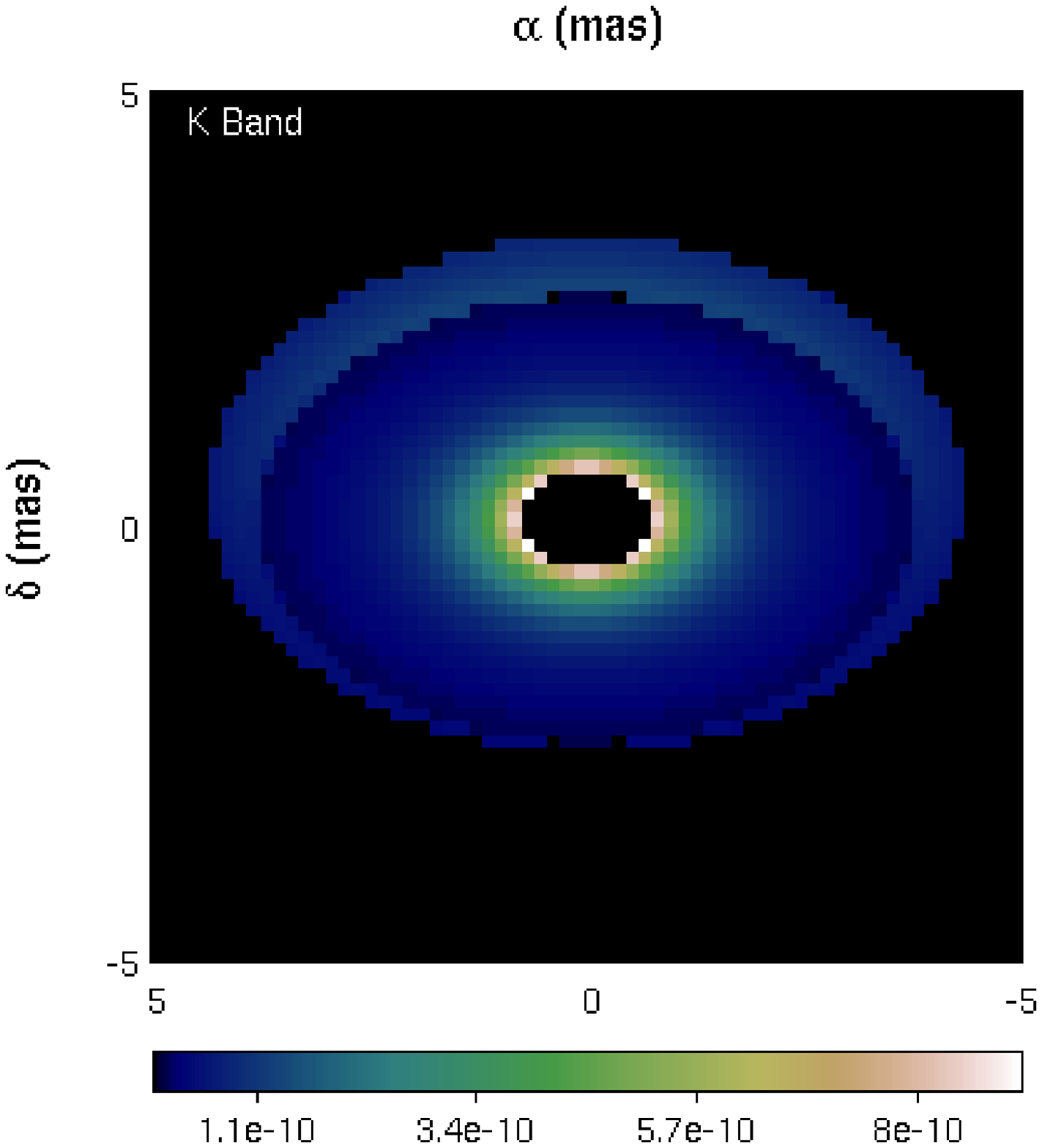}
& 
\includegraphics[width=0.42\textwidth]{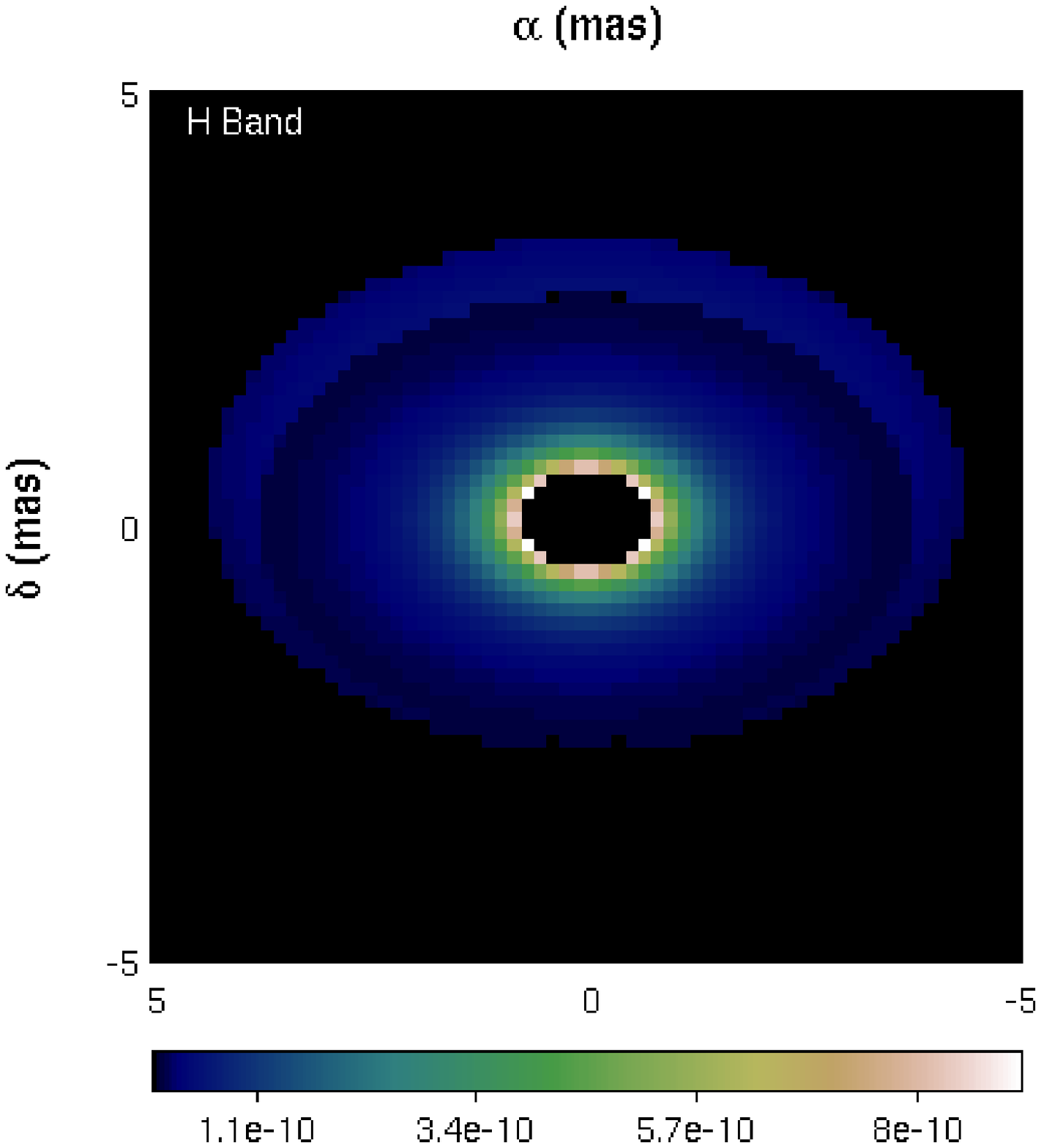}\\
\end{tabular}
\caption{\label{fig:image}  Model images  of the  circumstellar matter
  surrounding  HD~163296,  in  the   case  where  the  inner  disk  is
  consists  of refractory  iron grains  and  a rim  (outer ring).  The
  colors represent the flux in W.s/m$^{2}$.}  
\end{figure*}

\subsubsection{An inner dusty disk ?} 


Inside the silicate sublimation  radius, not only gas but also
  more refractory grains can exist  and contribute to the observed NIR
  emission.  In this section, we  explore the possibility that a layer
  of refractory grains, extending between an inner and an outer radius
  (inside the rim), accounts for the interferometric and photometric 
  observations of  HD~163296.  We assume  that the layer  is optically
  thin  in  the  vertical   direction  with  surface  density  profile
  $\Sigma_{\rm{dust}}$ and that its vertical optical depth is proportional to
$1/r$, where $r$ is the distance from the star \citep{dalessio99}. 
  We compute grain  temperatures and emissivity in the  H and K bands,
  and vary the  inner and outer radii  (R$_{\rm{in}}$   and
  R$_{\rm{out}}$, respectively) as well as the optical depth of
the layer.  We then  computed the emission, visibilities, and closure
  phases for  models that include this  inner layer, the  star, and the
  rim.  We consider separately three grain species known to be
  refractory:  iron, graphite,  and corundum  (aluminium  oxides).  The
  iron and  corundum grain cross-sections are computed  for spherical
  grains     from    the     optical     constants    tabulated     in
  \cite{pollack94}, \cite{koike95}, and \cite{begemann97}.  We use the graphite cross
sections                         tabulated                        by
  B.~Draine\footnotemark{}\footnotetext{$\textrm{http://www.astro.princeton.edu/~draine/dust/dust.diel.html}$},
  based on  the optical  constants of \cite{laor93}.   For relatively
  small  grains,  the  radial  temperature  profile  follows  $\propto
  r^{-0.4}$ (Fig.~\ref{dust}, left).  For comparison, we also show the
  temperature  profile   of  very   large  grains  (which   behave  as
  blackbodies),  which have flat  opacity from  the UV  to the  IR ($T
  \propto r^{-0.5}$).\\ 
  These models are very simple, but probably not unrealistic.
  The  strongest approximation concerns the dust temperature, which we
  compute by assuming that each grain is heated by the stellar radiation,
  and  attenuated  by  an  average optical  depth  $\tau_{star}\propto
  0.25$. In  fact,  the  attenuation is  not
  constant, since the optical path toward any individual grain varies not
  only  with $r$  but  also with  the  incident angle  of the  stellar
  radiation.  
  Once the  temperature of the grains  is known, then  the emission is
  computed   at   all   NIR   wavelengths  using   a   ray-tracing
  algorithm.  As discussed in the following, a rim caused by silicate
condensation can form  in the low density region that we propose and its
  properties  may also  affected  by  refractory
  grains  in the  inner disk,  which absorb  a small  fraction  of the
  stellar  radiation.  However, a  self-consistent calculation  of the
  rim properties is beyond the purpose of this paper.  In this section, we 
model the rim  following \cite{isella05} and \cite{isella06}, assuming
  micron-size silicates,  an evaporation temperature  of $\sim$1250~K,
  and an effective stellar luminosity of 75\%~L$_\star$ to account for the shielding by
  the inner disk.  The rim  radius is about 0.45~AU and its effective
  temperature is about 1000~K.  

\begin{figure*}[t]
\centering
\begin{tabular}{ccc}

\includegraphics[width=0.32\textwidth]{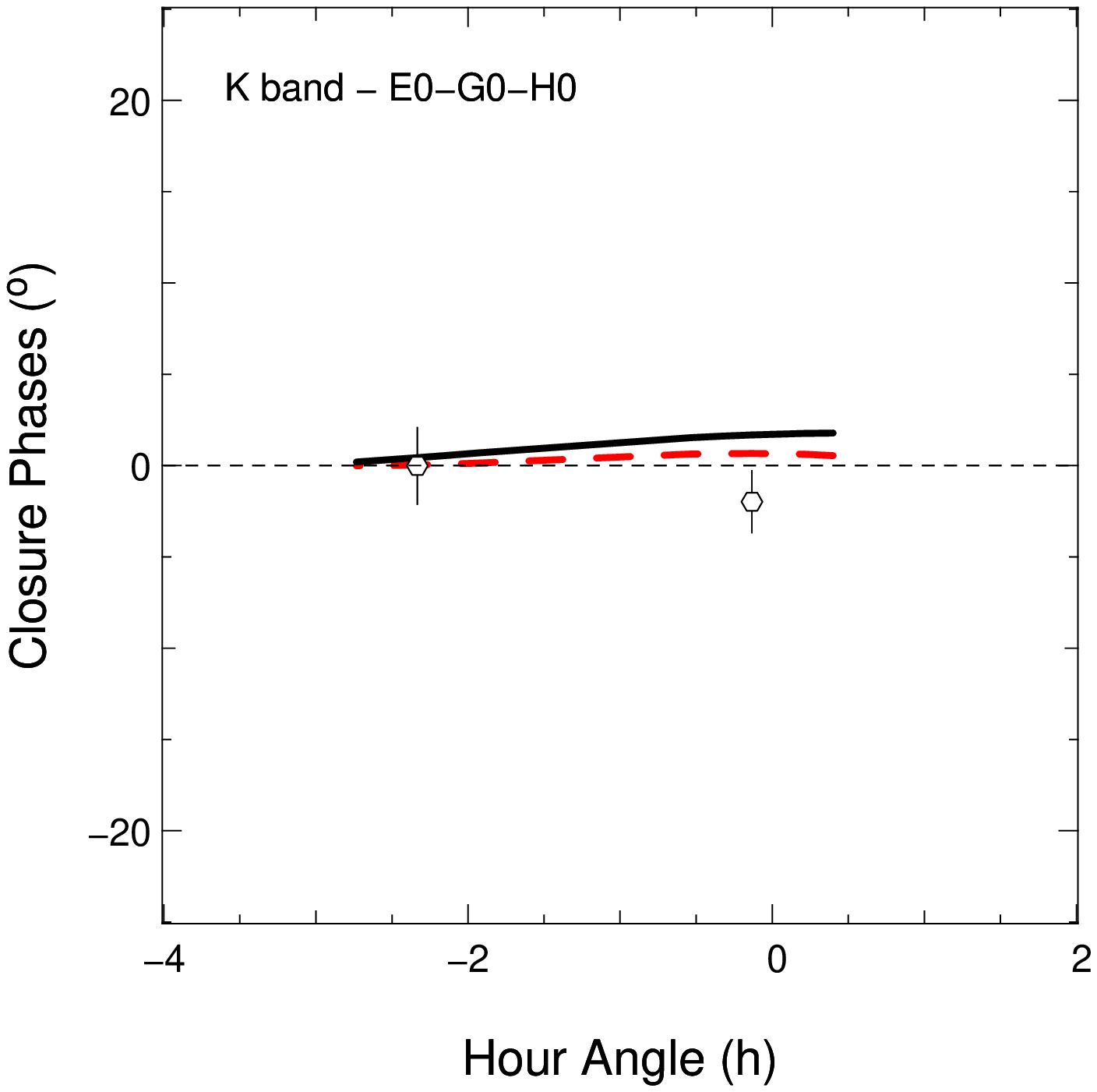}& 
\includegraphics[width=0.315\textwidth]{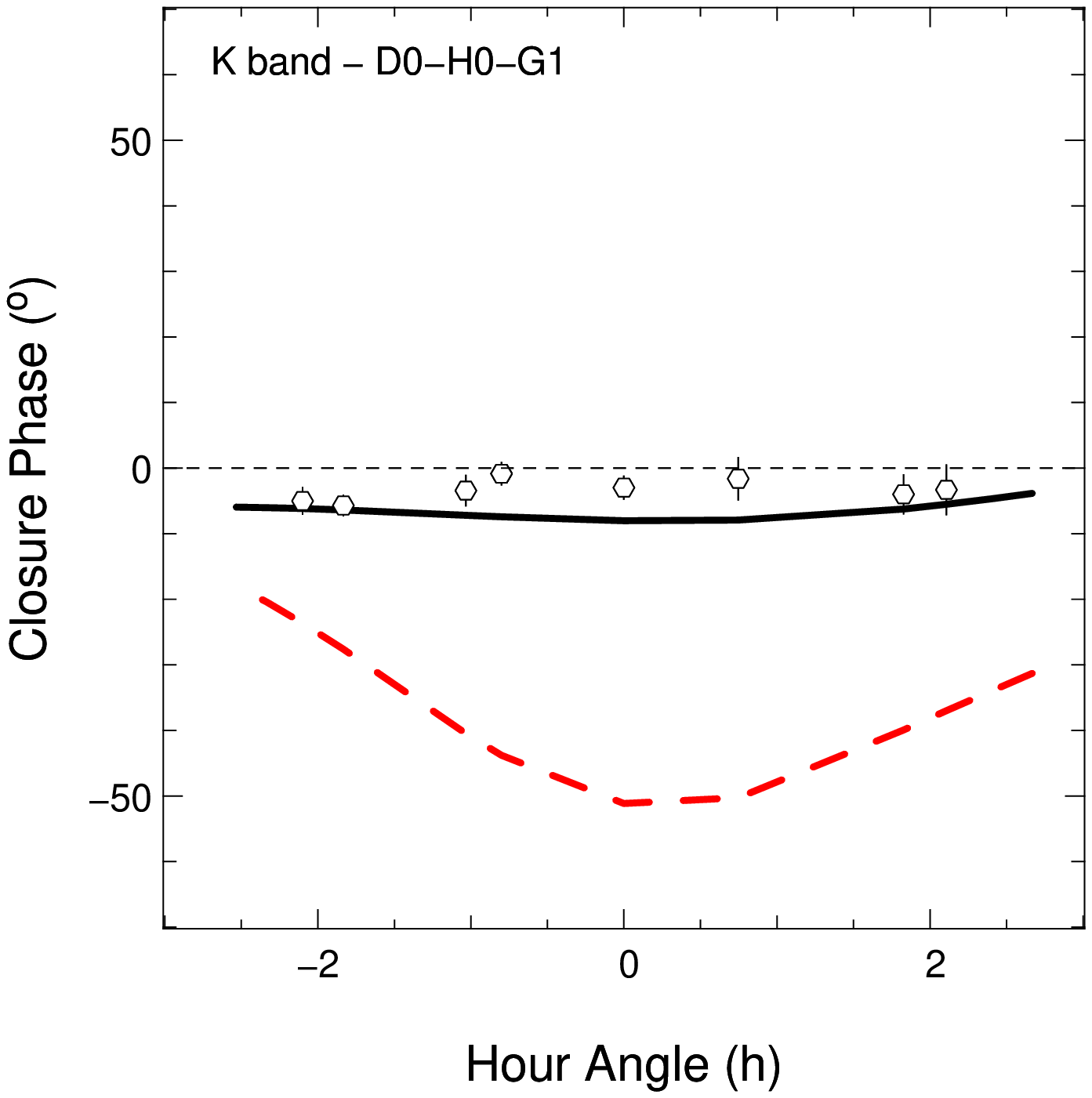}
&
\includegraphics[width=0.32\textwidth]{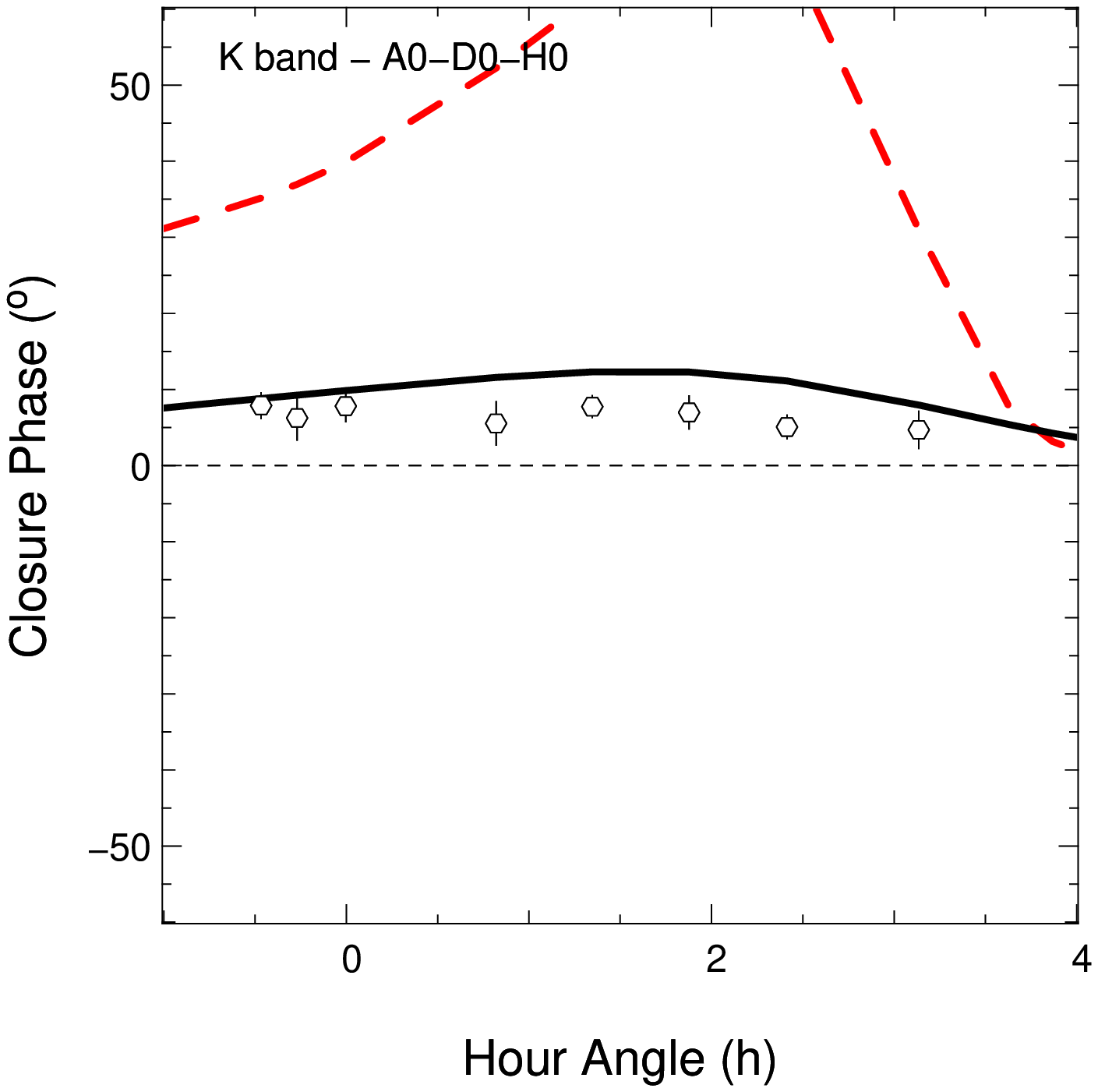}\\
\includegraphics[width=0.32\textwidth]{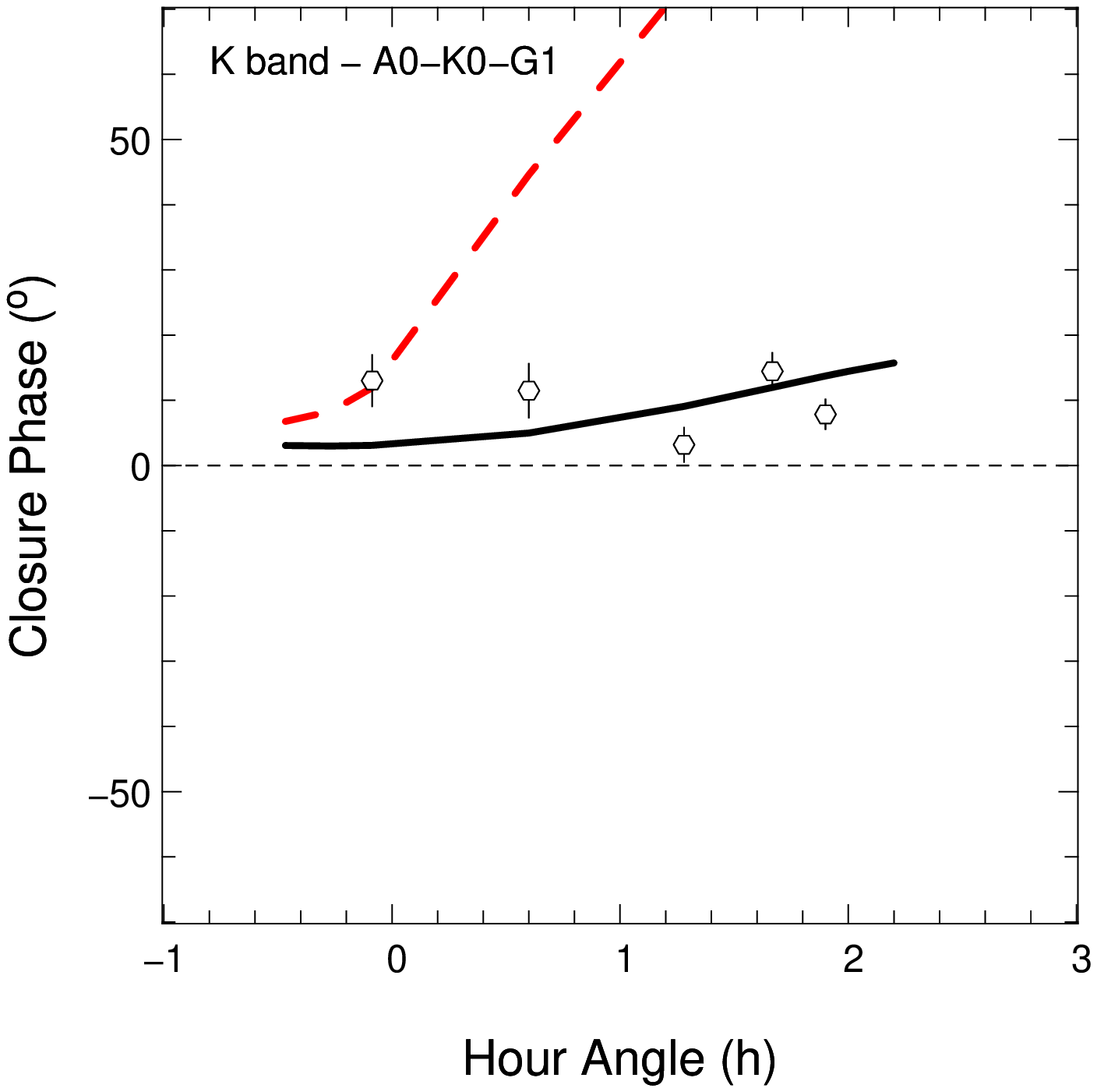}
& 
\includegraphics[width=0.315\textwidth]{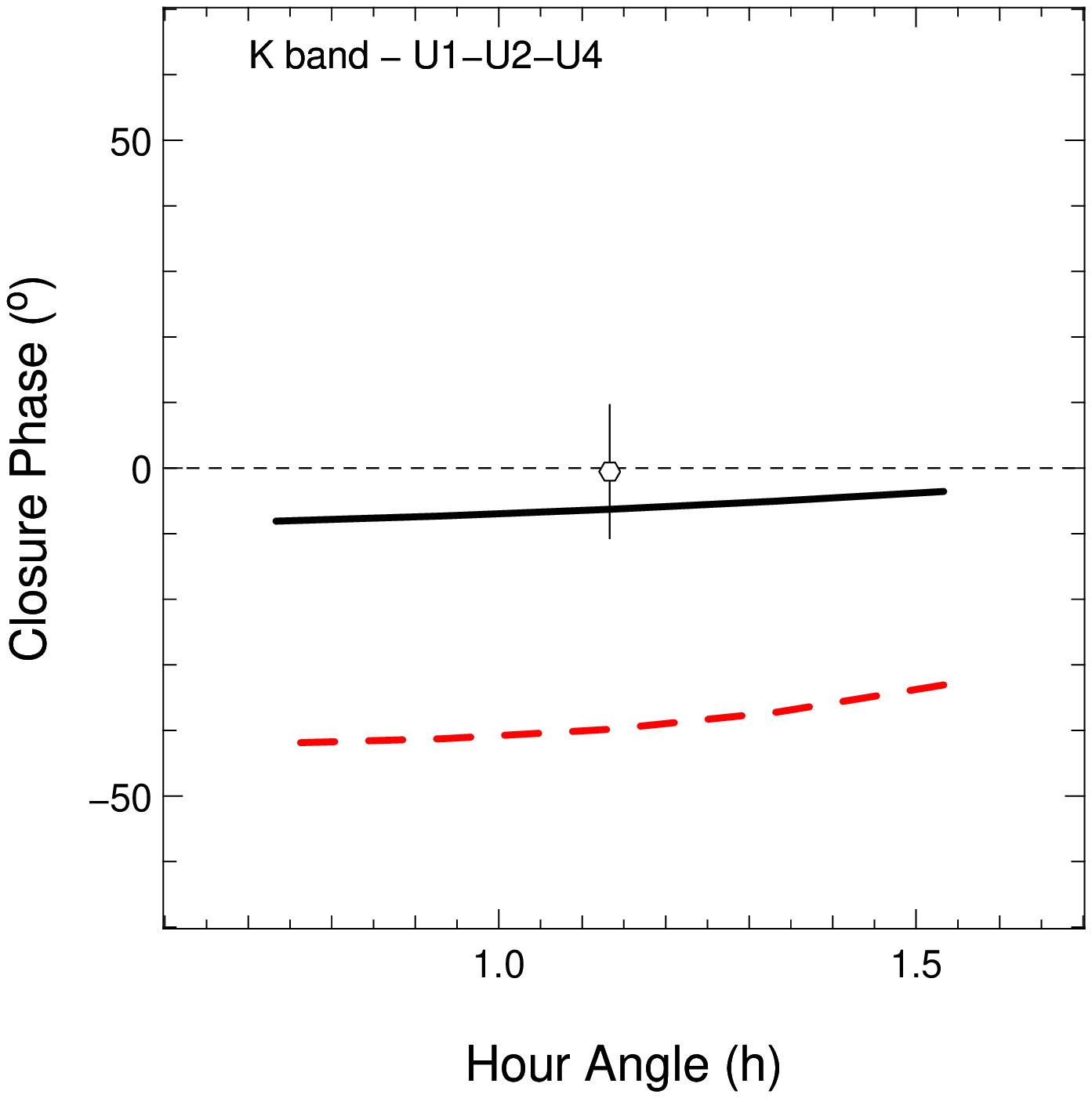}&
\includegraphics[width=0.325\textwidth]{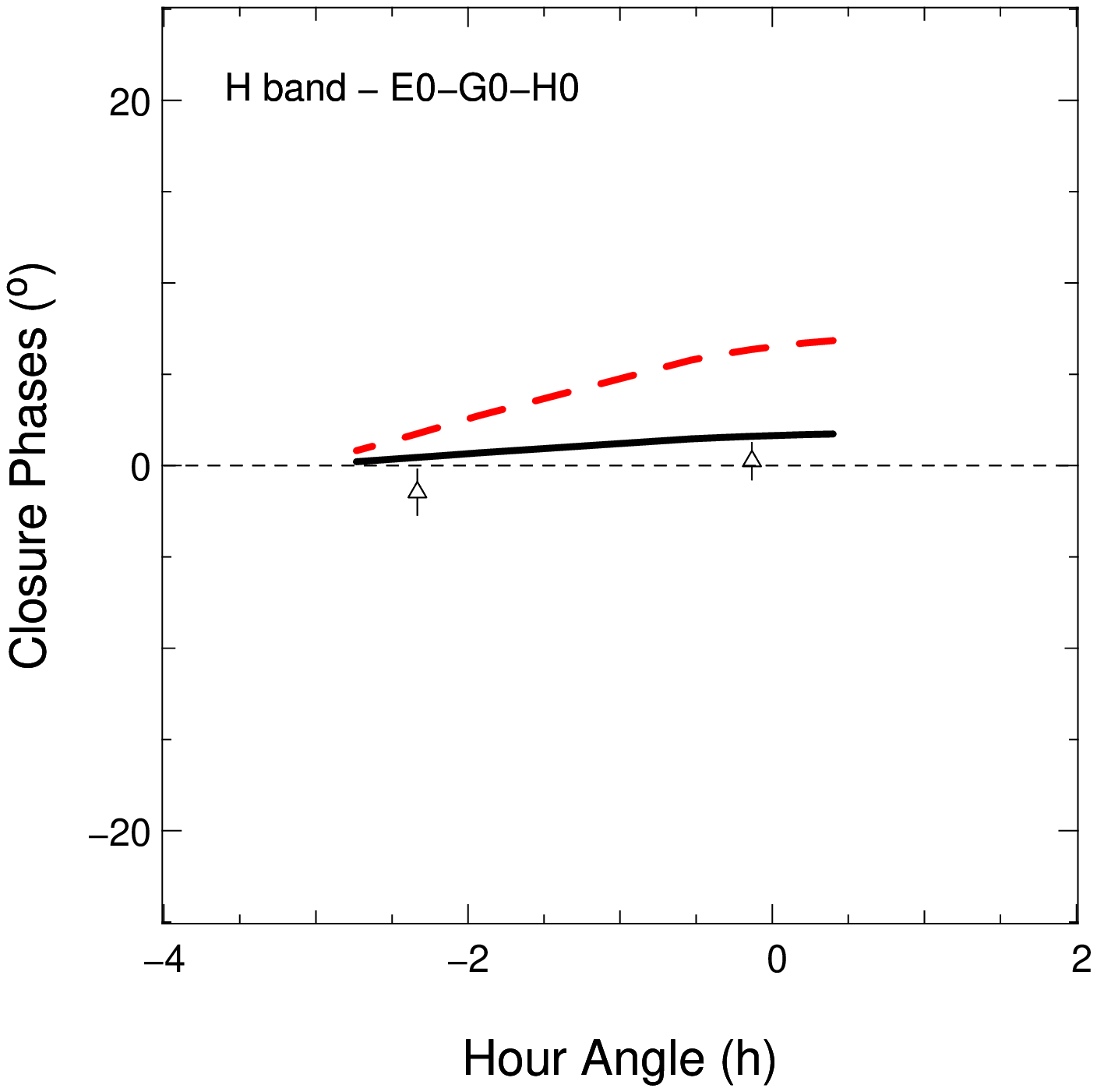}
\\
\includegraphics[width=0.32\textwidth]{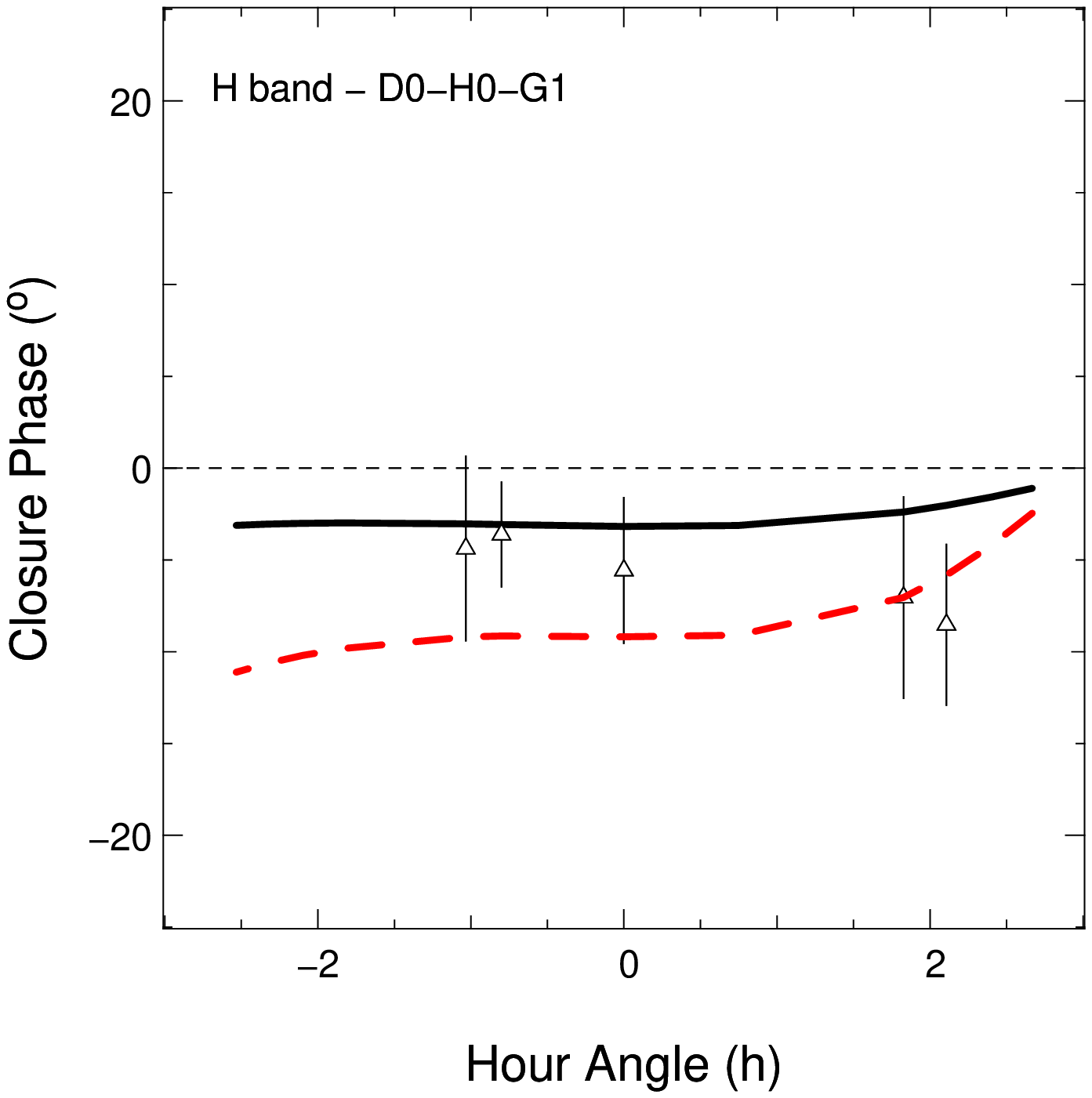}& 
\includegraphics[width=0.32\textwidth]{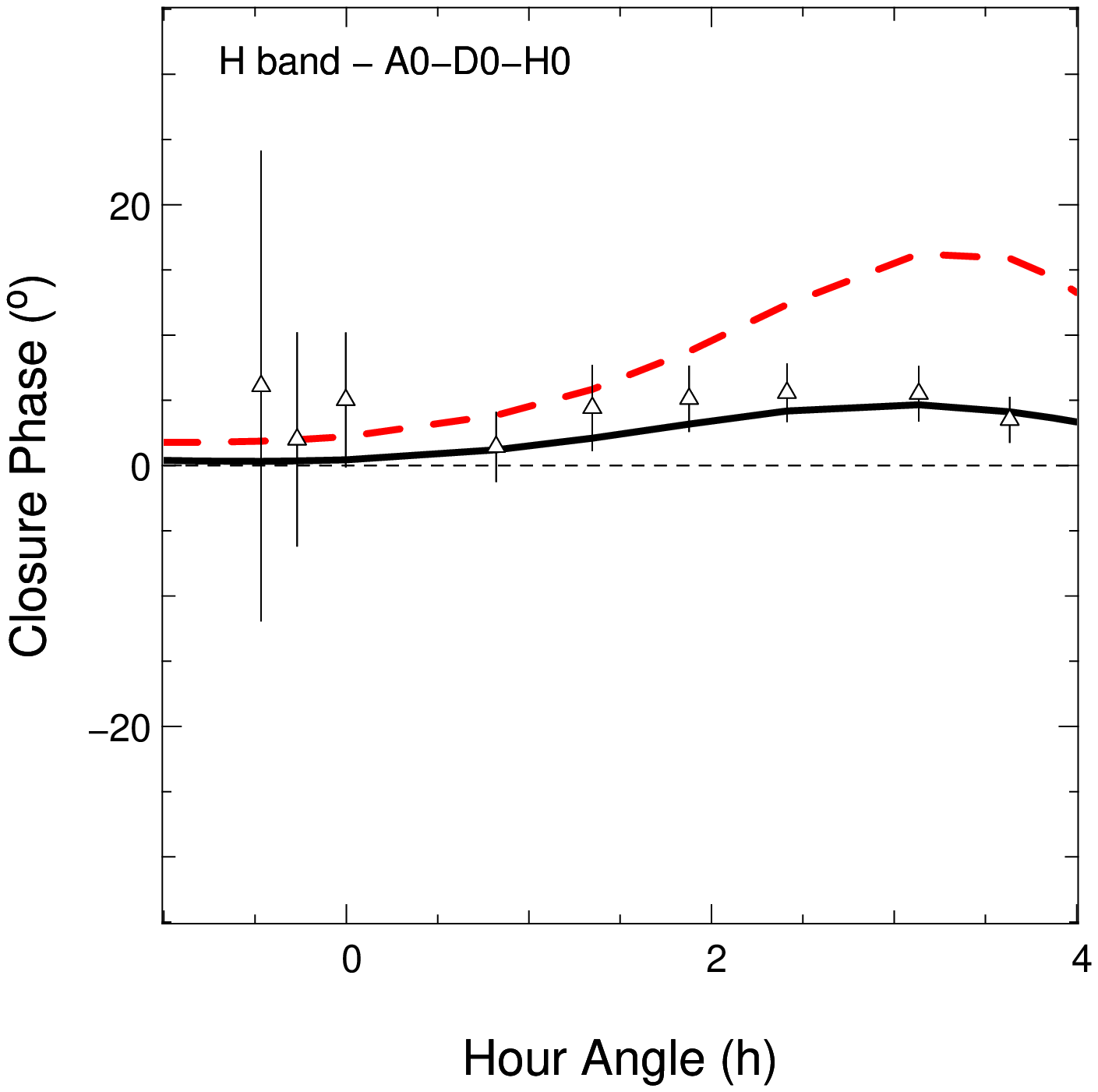}
& 
\includegraphics[width=0.32\textwidth]{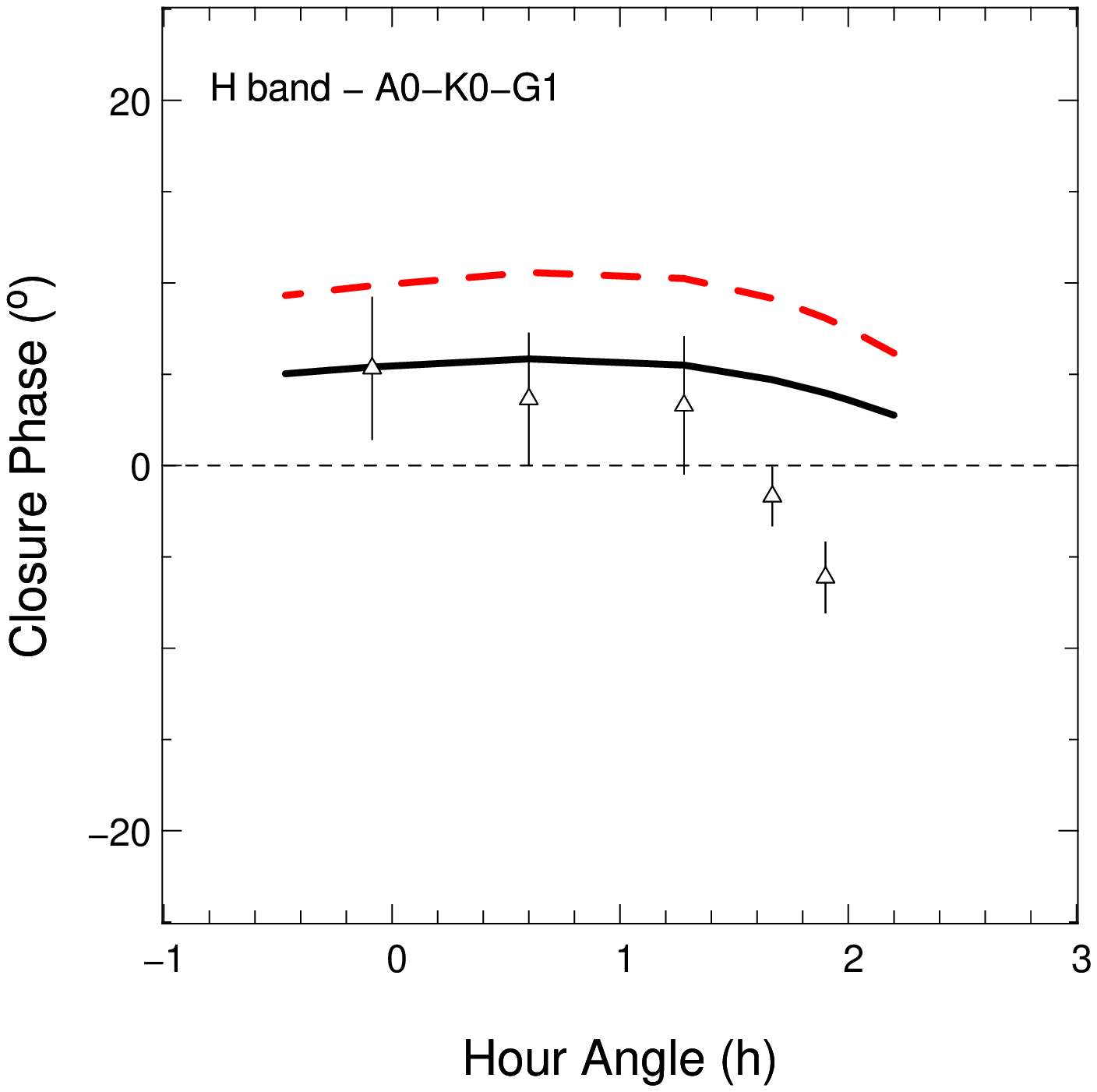}\\

\end{tabular}
\caption{\label{cpmod}  Closure  phases versus  hour  angle.  In  each
  panel, the predictions of the disk rim model (dashed red lines) and our
  model that  includes iron grains (full black  line) are overplotted.
  The  corresponding telescope configuration  and wavelength  band are
  indicated  in the  upper, left  corners. The  $K$ band  and  $H$ band
  measurements are plotted with circles and triangles, respectively.}  
\end{figure*}

We obtain a good fit to the SED, as shown in Fig.~\ref{dust} (right),
for $R_{\rm{in}}=0.10$~AU and $R_{\rm{out}}=0.45$~AU, and an optical
depth in the $H$ band at $R_{\rm{in}}$ of 0.31, 0.2, and 0.25 for iron,
graphite, and corundum, respectively.    Table~\ref{tab:grains} indicates the grains properties used in these models. The total mass within $0.45$~AU, which
is always   very   low   ($\leq  10^{-6}$~M$_{\oplus}$),   is   also
reported. Although we have not tried to constrain the grain parameters
in any detail, we note that very large refractory grains do not provide an equally
good fit to the SED, as the  ratio of the $H$ to the $K$ emission is
always too high.  \\ These models successfully reproduce the interferometric
data,  in terms  of both  visibilities and  CP.  Since  all  three models
produce similar 
results,  we present  the  case for  iron grains.  Figure~\ref{fig:bf}
shows the broad-band visibilities  compared with the predictions of such
a  model.  Figure~\ref{cpmod} presents  the broadband  closure phases
plotted  versus  hour  angle   and  the  model  predictions
calculated  for   each  telescope  configuration.    We  overplot  the
predictions of the rim-alone models (red, dashed line) to show how
adding the optically  thin dusty, inner disk emission  smoothes out the
asymmetry    induced   by    the    rim   by    the   correct    amount.
Table~\ref{tab:bestmodels}  summarizes  the  structural parameters  of
this model, as  inferred from SED and interferometric  data fitting, and
Fig.~\ref{fig:image} shows the corresponding $H$ and $K$ band images.

\begin{table}[tb]
\centering
\caption{\label{tab:grains}  Refractory  dust  model parameters.  For
  each species,  the optical  depths in $H$  and $K$ bands  at 0.10~AU
  ($\rm{R_{\rm{in}}}$) are reported as well as the minimum and maximum
  grain sizes  considered in the calculations. 
  The  resulting surface  density $\Sigma_{\rm{dust}}$  and  the total
  mass of dust grains are also given.}  
\begin{tabular}{c|cccccc}
\hline
\hline
Species & $\tau_{\rm{K, R_{\rm{in}}}}$& $\tau_{\rm{H}, R_{\rm{in}}}$ &
a$_{\rm{min}}$ & a$_{\rm{max}}$ & $\Sigma_{\rm{dust}}$ & M$_{\rm{dust}}$\\
 & & & [$\mu$m] & [$\mu$m]& [g cm$^{-2}$] & [M$_{\oplus}$]\\
\hline
\hline
Iron &  0.25 & 0.31 & 0.2 & 2 & 2.3~10$^{-4}$& 9.5~10$^{-7}$\\ 
Graphite & 0.14 & 0.20 & 0.05 & 0.5 & 2.1~10$^{-5}$ & 8.7~10$^{-8}$ \\
Corundum & 0.16 & 0.25 & 0.7 & 5 & 3.2~10$^{-4}$ & 1.3~10$^{-6}$ \\
\hline
\end{tabular}
\end{table}

\begin{table}[tb]
\centering
\caption{\label{tab:bestmodels}
  Parameters of  our model with iron grains distributed between R$_{\rm{in}}$ and
  R$_{\rm{rim}}$, the location of the disk rim.  For  both bands, the
  ratio  of the  stellar, rim, and inner disk contributions
  (F$_{*}$, F$_{\rm{rim}}$,  and F respectively)  to the total  flux in
  the \textit{model} (F$_{\rm{tot}}$) are reported.} 
\begin{tabular}{c|cccccccc}
\hline
\hline
Wavelength   &    R$_{\rm{rim}}$&   R$_{\rm{in}}$   &   T$_{\rm{in}}$&
F$_{*}$/F$_{\rm{tot}}$&
F$_{\rm{rim}}$/F$_{\rm{tot}}$&F/F$_{\rm{tot}}$\\ 
 & [AU]& [AU] & [K]& [\%] & [\%]& [\%]\\ 
\hline
\hline
$K$ band & 0.45 & 0.10 &  2100 & 14 & 36 & 50\\ 
$H$ band & ''& ''&  ''&  30 & 16 & 54\\
\hline
\hline
\end{tabular}
\end{table}

\vskip 0.1cm
The gas density of the inner disk can be derived from the values of
$\Sigma_{\rm{dust}}$, once  the abundances of iron,  carbon, and aluminium
in the  solid species are known.  Assuming, for example,  that 50\% of
the iron is in grains, the gas surface density at
0.10~AU will  be 0.2 g/cm$^2$ (0.02  g/cm$^2$ if 30\% of  carbon is in
graphite, or 1.3  g/cm$^2$ if all aluminium is  in corundum).  The gas
density can be higher if a lower fraction of the metals is condensed.  
However, it seems likely, from the considerations of Sect.~4.5.1,
  that the gas density cannot be  too high. A density only a few times
  higher than the above lower limits is inferred by
  the properties of the rim in HD~163296.  In particular, both the large
  rim radius and its low  effective temperature can be reproduced in a
  low density disk by silicates of micron size, as inferred in several
  HAe  stars \citep{isella06}, and  do not  require very  small grains.
  Assuming that  the rim  is produced by the evaporation of  silicates and
  that all the silicon is in olivine of $\sim 1\mu$m size, we analyzed the
  rim  properties  in disks  of  increasing  (but  still low)  surface
  density.  For $\Sigma= 5(r/0.1AU)^{-1}$ g/cm$^2$, the 
  \textit{vertical} optical depth to the stellar radiation is $\sim$10
  for a gas surface density of 1 g/cm$^2$ at $R_{\rm rim}$, 
  large enough to  allow the formation of an  optically thick rim that
  can be  modelled following \cite{isella05}.  The low  gas density, and
  the correspondingly low evaporation  temperature, moves the rim radius
  further from the star.  Assuming a reasonable scale height of 
$10^{-3}-10^{-2}$~AU, the gas density is $\sim 10^{-11}$~g/cm$^3$, and
the silicate evaporation temperature of the order of 1150~K.  The
corresponding rim radius is $\sim$0.4-0.5~AU, and its effective
temperature is about 1000-1100~K, as required to fit the HD~163296
SED and interferometric data.  \\ 

  The only difficulty in assuming that the inner disk
  emission originates in grains within the silicate sublimation
radius, is the need for  them to survive at temperatures much higher
  than is generally assumed.  The three types of grains that we
examined  reach  temperatures   of  2100-2300~K  at  $\sim  0.10$~AU
  (Fig.~\ref{dust},  left).   While similar  values  are possible  for
  graphite \citep{krugel03},  they are too high for both iron and corundum
  in the pressure of our inner disk \citep{pollack94,posch03, kama09}.
  However,  there  is room  for  discussion  \citep{najita09}, as  the
  balance between gas and dust in the conditions of the inner disk should be 
  reconsidered in detail \citep{duschl99}. We emphasize that only
  a small amount of refractory grains need to survive these high
temperatures, \textit{i.e.}, probably a minor fraction of the original
  population.   The   grains  that  we  have   considered  are  likely
  candidates,  but if other,  more refractory  species can  form, they
  would certainly fit the observations equally well.  

 \section{Summary and conclusions}
 This paper had discussed the largest set of NIR interferometric data
 collected so far for a young star.  HD~163296 has a well
 studied disk at large spatial scale, which motivated our interpretation of
 the  NIR  interferometry  using  a  star +  inner  disk  model.   Both
 interferometric  and  photometric data  can  be  accounted  for by  an
 inclined  disk with  a low  density inner  region.  The  NIR continuum
 emission is then not dominated by the thermal emission
 from  the  dusty  disk  rim  located  at  the  sublimation  radius  of
 astronomical silicates, but by the additional optically thin component located inside. 
 This component emits about 32\% of the stellar luminosity, but
 54\% (50\%) of the observed radiation in $H$ ($K$) band.  Because of the
 silicate condensation  and the strong  increase in opacity,  the disk
 rim forms in this low-density  region at $\sim 0.45$~AU from the star
 and  emits  about 16\%  (36\%)  of the  $H$  ($K$)  band flux.   The
 combination  of  the unresolved  stellar  radiation,  the smooth  and
 point-symmetric emission of the inner disk region, and the skewed disk
 rim emission  can successfully explain the  visibilities and non-zero
 closure phases measured in both bands.  

The nature of the emission in the inner disk remains a matter of 
discussion.  We argue against gas being  mainly responsible
  for this continuum emission.  A dense, cold disk, as expected
for viscous accretion models (Muzerolle et al. 2004), would
  produce strong molecular lines  that are not seen in high-resolution
  spectra. Non-LTE tenuous gas layers, in an atomic state if only heating
  by the  star is assumed, or  fully ionized by  additional sources of
  energy, cannot account  for the observed  properties of the  NIR continuum.
However, we  emphasize that  self-consistent models  of dust-free gaseous
  disks are  not currently  available, but are  needed to  exploit the
  full potential of the interferometric observations.  

We suggest instead that a small fraction of refractory grains survive very
close to the star.  We propose models for the optically
thin  emission  of the  innermost  region,  using  
various kinds of refractory grains,  distributed from  0.10~AU  to 0.45~AU.
The dust  surface density provides  only a lower limit  to the
  gas surface density, as we do know neither the exact nature of the
grains nor  their abundance.   However, we find  that a  low density
  region is consistent with the location and properties of the rim, as
  condensation of silicates will occur,  and with the lack of molecular
  features in the  spectrum of HD~163296.  We expect the gas
  in the inner disk indeed to be mostly atomic, in non-LTE, and although its continuum
  emission will  be weak, hydrogen  lines can be strong.   The models
  used in Sect.~4.5.2 to argue  for the presence of refractory grains,
  make  a  number  of  crude  assumptions,  and  improved  models  that
  self-consistently compute the grain temperature and emission 
  in  the thin  disk as  well as  the properties  of the  rim  are being
  developed.   Our  study  indicates   that  the  inner  region  of
HD~163296 is quite 
empty, with a  very low surface density that  is inconsistent with a
  dense  accretion  disk.   For   comparison,  a  surface  density  of
  1~g/cm$^2$ at $\sim$0.10~AU corresponds, in a standard accretion disk
   ($\alpha$=0.01),     to     an     accretion     rate     of     $\sim
   10^{-11}$~M$_{\odot}$yr$^{-1}$, much lower  than typical values for
   Herbig~Ae stars  \citep{garcialopez06}.  With  our data, we  do not
   constrain the outer radius of this low density region, which can be
   larger than  0.5~AU. However,  we know that  at large radii  the HD
   163296  disk is  massive  and  dense, as  shown  by the  millimeter
   interferometric observations of \cite{isella07}.  
 It seems likely that HD~163296 has a dense disk with an inner cavity,
 and that  we observe it  just before it  reaches the transition
 disk phase, as suggested by \cite{sitko08}. 

 \vskip 0.1cm

 It  is  fair  to  emphasize  that  we  make  no  claim  that  our
   interpretation is unique. As stated at the beginning of Sect.~4, we
   assumed that the NIR emission of HD 163296, at spatial scales 
   of less than 0.5~AU, is dominated by the emission of a circumstellar disk.
   Moreover, we interpreted the non-zero values of the CP as evidence
   of the  asymmetric emission  of a disk  rim.  While  the properties
   that we derived for the  smooth inner emission are probably robust,
   the existence  of the  rim is  less so. In  particular, the  lack of
   visibility bounces  at large baselines argues  against the presence
   of a rim.   In this case, the observed  closure phases may possibly
   be caused by any asymmetric brightness distribution, 
such as a symmetric flared disk with a stellar contribution
   that is off-centered by a few percent of the inner disk radius with
   respect to the disk \citep{malbet01}, by a hot spot on the disk, or by a
   density  discontinuity.   Only  a larger  \textit{(u,v)}  coverage,
   providing access  to more details  of the morphology could  solve this
   ambiguity.\\ 

 Our interpretation  of the smooth,  inner emission as  originating in
 refractory grains requires their survival at very high 
 temperatures ($\sim 2100-2300$ K), much higher than expected, even for
 the most refractory  grains, at the pressure of  the low-density inner
 disk \citep{pollack94,  kama09}.  However, clearing  the innermost
 regions and optically thin  emission from left-over refractory grains
 could be  a common phenomenon  in Herbig~Ae stars. Their  presence within
 the first  few tenths of an  AU is a promising  interpretation of the
 observed depletions in refractory dust species - such as iron - 
 in   jets   of   young    stars   that   are   launched   from   this
 region~\citep{nisini05,  podio06}.  Similar  interferometric studies,
 with  a large  number  of measurements  in  various wavelength  bands
 simultaneously, should be performed for  a large sample of stars.  A higher
 level of complexity in models is  also needed to account for both the
 dust and  the gas  emission in a  self-consistent way.   Finally, the
 advent  of  the next-generation of imaging  instruments will  hopefully
 provide unambiguous constraints on these complex environments.


\begin{acknowledgements}
  We  acknowledge   fundings  from  CNRS  and   INAF  (grant  ASI-INAF
  I/016/07/0).  This  work was in  part performed under  contract with
  the  Jet Propulsion  Laboratory  (JPL) funded  by  NASA through  the
  Michelson  Fellowship  Program.  JPL  is  managed for  NASA  by  the
  California Institute of Technology.   We thank A.~Sargent and T.~Ray
  for hosting part of this research,  and the VLTI team at Paranal for
  the help in obtaining these data. We thank J.D.~Monnier who provided
  with  the  IOTA  data,   and  M.~Sitko  for  discussions  about  the
  photometric  measurements.  We are  greatful to  A.~Crida, M.~Desort
  and  S.~Renard   for  fruitful  discussions.    We  acknowledge  the
  anonymous referee for his comments  that improved the clarity of the
  paper. 
\end{acknowledgements}

\newpage

\bibliographystyle{aa}
\bibliography{bib_mwc275}

\end{document}